\documentclass[fleqn,usenatbib]{mnras}

\usepackage{newtxtext,newtxmath}

\usepackage[T1]{fontenc}
\usepackage{ae,aecompl}

\usepackage[pdftex]{graphicx}	
\usepackage{amsmath}	
\usepackage{amssymb}	

\usepackage{epsfig}
\usepackage{natbib}
\usepackage{threeparttable} 
\usepackage{xspace}
\usepackage{color}
\usepackage{epstopdf}
\usepackage{gensymb}
\usepackage{booktabs}
\usepackage{times}
\usepackage[utf8x]{inputenc}
\usepackage{array}

\newcommand{\athena}{\textit{Athena}\xspace}

\newcommand{\chan}{\textit{Chandra}\xspace}

\newcommand{\rxte}{\textit{RXTE}\xspace}
\newcommand{\suzaku}{\textit{Suzaku}\xspace}
\newcommand{\asca}{\textit{ASCA}\xspace}

\newcommand{\xmm}{\textit{XMM-Newton}\xspace}

\newcommand{\gaia}{\textit{Gaia}\xspace}

\newcommand{\Msun}{\mathrm{M}_{\odot}}

\newcommand{\nh}{\mathrm{cm}^{-2}}

\newcommand{\mxb}{MXB~1659--29}
\newcommand{\exo}{EXO 0748--676}

\newcommand{\aql}{Aql X-1}
\newcommand{\ks}{KS~1731--260}
\newcommand{\cen}{Cen X-4}

\newcommand{\uu}{4U 1608--52}
\newcommand{\xte}{XTE J1701--462}

\title[]{On obtaining neutron-star mass and radius constraints from quiescent low-mass X-ray binaries in the Galactic plane}

\author[A. Marino et al.]{
Alessio Marino$^{1}$\thanks{E-mail: alessio.marino@unipa.it},
N. Degenaar$^{2}$,
T. Di Salvo$^{1}$,
R. Wijnands$^{2}$,
L. Burderi$^{3}$
and R. Iaria$^{1}$
\\
$^{1}$Universit\`a degli Studi di
  Palermo, Dipartimento di Fisica e Chimica, via Archirafi 36 - 90123 Palermo, Italy\\
$^{2}$Anton Pannekoek Institute for Astronomy, University of Amsterdam, Postbus 94249, 1090 GE Amsterdam The Netherlands \\
$^{3}$Universit\`a degli Studi di
  Cagliari, Dipartimento di Fisica, SP Monserrato-Sestu km 0.7, I-09042 Monserrato, Italy
}

\date{Accepted XXX. Received YYY; in original form ZZZ}

\pubyear{2018}

\hypersetup{draft}
\begin{document}
\label{firstpage}
\pagerange{\pageref{firstpage}--\pageref{lastpage}}
\maketitle

\begin{abstract}
X-ray spectral analysis of quiescent low-mass X-ray binaries (LMXBs) has been one of the most common tools to measure the radius of neutron stars (NSs) for over a decade. So far, this method has been mainly applied to NSs in globular clusters, primarily because of their well-constrained distances. Here, we study \chan\ data of seven transient  LMXBs in the Galactic plane in quiescence to investigate the potential of constraining the radius (and mass) of the NSs inhabiting these systems. We find that only two of these objects had X-ray spectra of sufficient quality to obtain reasonable constraints on the radius, with the most stringent being an upper limit of $R\lesssim$14.5~km for EXO 0748--676 (for assumed ranges for mass and distance). Using these seven sources, we also investigate systematic biases on the mass/radius determination; for Aql X-1 we find that omitting a power-law spectral component when it does not seem to be required by the data, results in peculiar trends in the obtained radius with changing mass and distance. For \exo\ we find that a slight variation in the lower limit of the energy range chosen for the fit leads to systematically different masses and radii. Finally, we simulated \athena\ spectra and found that some of the biases can be lifted when higher quality spectra are available and that, in general, the search for constraints on the equation of state of ultra-dense matter via NS radius and mass measurements may receive a considerable boost in the future.
\end{abstract}




\section{Introduction}\label{sec:intro}
Matter in the interiors of neutron stars (NSs) is found in extreme conditions (for instance a density at least $\sim$2--5 times higher than the nuclear density), and hence requires a special equation of state (EoS) of ultra-dense matter to describe the composition and behaviour. The dense-matter EoS determines the mass ($M$) and radius ($R$) of a NS by means of the Tolman-Oppenheimer-Volkoff equations and results in specific $M-R$ curves for each EoS \citep[see][for a recent review]{Ozel2016}. Each EoS is characterized by a maximum mass, resulting from the fact that all EoSs have a maximum central density beyond which no stable configuration is possible. 

Dozens of possible $M-R$ relations have been developed in the past few decades \citep[see e.g.][]{Lattimer2012}, all allowed within our current understanding of NSs. The credibility of different EoSs to describe NS interiors cannot be tested with terrestrial experiments, because it is impossible to reach the required extreme densities and, at same time, the low temperatures relevant for NSs in any laboratory on Earth. This is therefore an astrophysical task: directly measuring masses and radii of NSs, or even only obtaining some constraints on these parameters, can tell us if a theoretical EoS is plausible or not. 

Unfortunately it is not easy to measure both the radius and the mass of a single NS. Whereas masses have been reliably determined for a number of radio pulsars through radio pulsar timing techniques \citep[e.g.][]{Ozel2016}, it is very challenging to determine the radii of these NSs \citep[see][for a recent discussion and outlook]{Watts2015}. Without a measurement of both $M$ and $R$ for the same NS, it is only the most extreme mass measurements, near and above $\sim 2~\Msun$, that start to put some interesting constraints on the dense-matter EoS \citep[e.g.][]{Demorest2010,Antoniadis2013,Hebeler2013,Fonseca2016,Fortin2017}. 

In the past decade, some relevant steps forward have been made in estimating the radii of NSs through the development of various techniques.
Nearly all of these rely on X-ray spectroscopy and the most promising ones involve analyzing the thermal emission emerging from the surface of the hot NSs in low mass X-ray binaries (LMXBs). This surface emission provides the means to directly measure their apparent angular size and extract information on the radius. However, to convert the radius observed in infinity to the true NS radius requires to take into account gravitational effects, so that the NS mass enters the equations as well; in practice a combined mass/radius pair is thus measured. If independent constraints can be obtained, for instance on the gravitational redshift or on the dynamical mass, the degeneracy can be broken and both $M$ and $R$ can be obtained \citep[e.g.][]{Ozel2006}.

The surface emission from NSs is often overwhelmed by other emission processes in LMXBs, most prominently the X-rays from the accretion flow, but it is visible under a number of circumstances. For instance, $M$ and $R$ measurements can in principle be obtained from observations of thermonuclear X-ray bursts (simply X-ray bursts hereafter), because the NS surface then briefly outshines the accretion flow \citep[e.g.][]{Ebisuzaki1987,Vanparadijs1990,Lewin1993,Zamfir2012,Ozel2016,Nattila2016}. Furthermore, some NSs channel the accretion flow along their magnetic field lines creating visible surface hotspots; the resulting X-ray pulse profiles can be a powerful tool to extract information on the NS mass and radius \citep[e.g.][]{Pavlov97,Nath2002,Poutanen2003,Leahy2008,Lo2013,Miller2015,Bogdanov2016}.

A third method, which is the focus of the present work, consists of analyzing the quiescent thermal emission from transient LMXBs; the heat radiation released from these systems during the phase in which the accretion has (nearly) switched off. In this quiescent state these sources are $\sim$4--5 orders of magnitude fainter than during their active accretion episodes, with an X-ray luminosity of $\approx$ 10$^{32}$--10$^{33}$~erg$~$s$^{-1}$. The heat radiated from the surface is thought to be generated in nuclear reactions \citep[e.g.][]{Haensel2003} that occur in the crust during accretion phases (\citet{Brown1998}; see \citet{Wijnands2017} for an observational review).

The spectrum emerging from a NS is expected to be a blackbody reprocessed and modulated by the interactions with the stellar atmosphere. When a NS is in quiescence this thin layer is expected to be made of mostly H (or He if the companion is a degenerate star), because heavier elements should settle on short time scales.  To obtain reliable mass/radius measurements from studying the thermal surface emission of NS LMXBs, it is vital to model the atmosphere correctly \citep[e.g.][]{Zavlin1996,Rutledge1999,Suleimanov2011,Servillat2012,Nattila2015}. Several theoretical models have been developed to describe the quiescent thermal emission of the weakly magnetized ($B \lesssim 10^9$~G) NSs in LMXBs, assuming different compositions of the atmosphere \citep[e.g.][]{Zavlin1996,Heinke2006,Ho2009,Haakonsen2012}. 

Sometimes the emission from a quiescent LMXB (qLMXB hereafter) cannot be properly described by a simple NS atmosphere model due to the presence, often quite evident from the spectrum, of a hard emission tail that dominates over the thermal component at energies $\gtrsim$3~keV \citep[e.g.][]{Asai1998,Rutledge1999,Jonker2004,Cackett2011,Fridriksson,Chakrabarty2014,Parikh2017}.\footnote{The hard power-law tail in the quiescent spectra of NS LMXBs is often referred to as ``non-thermal'' because it deviates from the soft thermal spectral component that is ascribed to the NS surface emission. However, Bremsstrahlung emission from a boundary layer is a viable mechanism for the harder spectral component, so indicating it as non-thermal is strictly not correct; see \citet{Dangelo2015}.} 
The physical origin of this emission, usually modeled as a simple power law, is not clear but several possible explanations have been suggested in the past; these include residual accretion on to the NS surface and different emission processes related to the magnetic field of the NS \citep[e.g.][]{Campana1998_review, Degenaar2012, Chakrabarty2014, Dangelo2015, Wijnands2015}.

Inferring NS radii from their quiescent thermal emission is promising, but subject to various systematic uncertainties. First of all,  since the thermal emission essentially provides a measurement of the angular size, uncertainties in the source distance translate directly into uncertainties in the inferred radii. Moreover, apart from  the effect of technical issues such as pile-up \citep[e.g.][]{Bogdanov2016},
the (often unknown) atmosphere composition has a large impact on radius measurements of NS LMXBs \citep[e.g.][]{Servillat2012,Catuneanu2013,Heinke2014}. Biases are also introduced if the surface temperature of the NS is inhomogeneous \citep[][]{Elshamouty2016}; this could occur for instance if residual accretion takes place, which we know to happen in at least some sources \citep[e.g.][]{Cackett2010,Fridriksson,Dangelo2015}. Local channeling of heat from the interior along magnetic field lines can also potentially generate hotspots on the NS surface that bias the measurements \citep{Elsner1977,Ikhsanov2001,Lii2014,RoucoEscorial2017}. Finally, many NSs also spin very rapidly, which causes them to be oblate and their surface radiation to be Doppler boosted. However, these effects are smaller than the typical systematic uncertainties involved in EOS determinations and are therefore typically uncorrected for in this method \citep[e.g.][]{Steiner2013}.

\subsection{Motivation for the present work}
So far, mostly globular clusters have been exploited to measure NS radii and masses from the quiescent thermal emission. Among their often low interstellar extinction and the their relative abundance in these environments, a primary advantage of using NSs in globular clusters is their typically well-constrained distances opposed to field LMXBs \citep[e.g.][]{Heinke2003,Heinke2014,Webb2007,Guillot2013,Bogdanov2016,Steiner2017}.
Furthermore, qLMXBs in globular clusters often have purely thermal spectra, i.e. that do not require a hard emission component to be modeled, and show no temporal variations \citep[with some exceptions; e.g.][]{Heinke2003b,Degenaar2012_terzan5,Bahramian2015}. The absence of a power-law component in the X-ray spectra and lack of variability suggests that there is no ongoing accretion that could hamper obtaining reliable constraints on the NS parameters.

Nevertheless, there are also some drawbacks of using qLMXBs in globular clusters for this approach. In particular, most of the sources used in this kind of analysis have never exhibited an outburst and have faint companion stars, so that the composition of the NS atmosphere is not known. This is a concern because many LMXBs in globular clusters may have H-poor donor stars \citep[e.g.][]{Bahramian2014}; as mentioned above, uncertainties in the composition of the atmosphere have a huge effect on the determination of NS radii. 

Field LMXBs have the complication that their distances are often not accurately known \citep[current uncertainties are at the level of 10$\%$ or higher; e.g.][]{jonker2004_distances}, their quiescent spectra often contain a power-law spectral component \citep[e.g.][]{Jonker2004}, and their quiescent emission can vary irregularly on time scales of hours to years \citep[e.g.][]{Rutledge2002,Campana2003}. The latter two features are often taken as evidence that some residual accretion is ongoing \citep[e.g.][]{Cackett2010,Wijnands2015}.  
However, field LMXBs also offer certain advantages for NS mass and radius determinations. For instance, field LMXBs have typically undergone one or more accretion outbursts during which the companion type and/or composition of the accreted matter could be determined. Furthermore, during those outbursts the X-ray emission may be pulsed or X-ray bursts may be observed, which can provide independent measurements of $M$ and $R$. What further adds to field LMXBs as promising tests for the dense-matter EOS is that additional, indirect constraints on NS radii may come from measurements of the accretion flow properties (obtained during outbursts), such as the localization of the inner accretion disk via X-ray reflection spectroscopy \citep[e.g.][]{Bhattacharyya2010,Miller2013,Chiang2016,Ludlam2017, Matranga2017}, or studies of kHz and mHz quasi-periodic oscillations \citep[QPOs; e.g.][]{Miller1998,Vanstraaten2000,Barret2006,Stiele2016}. Although these measurements are typically not constraining by itself, there is some potential in combining them with more direct measurements for improved constraints on $M$ and $R$, and to cross-calibrate different methods \citep[e.g.][for a discussion]{Watts2016}. Another argument in favour of field LMXBs is that sometimes the companion stars are bright enough to allow for optical dynamical mass measurements during their quiescent states \citep[e.g.][]{Orosz1999,Elebert2009,Casares2010,Wang2013,Matasanchez2017}; although current constraints are not strong, these can possibly be improved with large upcoming facilities. Finally, it is important to explore how useful quiescent field LMXBs are for precise NS mass and radius measurements because new sensitive X-ray instruments, like the scheduled mission \athena\ (see Section~\ref{sec:simulations}), will not have the sub-arcsecond spatial resolution that is typically needed to resolve the globular cluster sources.

In this work we present a dedicated X-ray spectral analysis of seven LMXBs, located in the Galactic plane, during their quiescent states. This study is aimed, on one hand, to constrain masses and radii of the NSs inhabiting these systems, and on the other hand to investigate the main biases and weak spots of applying this method to field LMXBs.

\section{Sample and data selection}\label{sec:data}
For our study, we searched the literature for NS LMXBs in the field (i.e. globular cluster sources were not considered) that have been observed in quiescence with sensitive X-ray instruments (see Section~\ref{sec:reduction}). In total, quiescent data are available for $\sim$49 NS LMXBs \citep[see][for a recent overview]{Wijnands2017}. Detailed studies of the thermal emission of NS qLMXBs requires high sensitivity at soft X-ray energies ($\lesssim 3$~keV), and are therefore best carried out with the \chan\ or \xmm\ observatories. To refrain from possibly introducing an additional bias in our analysis related to instrument cross-calibration \citep[e.g.][]{Degenaar2011}, we here choose to focus solely on \chan\ data. 
A source suitable for our study needs to fulfill the following additional requirements: i) the quiescent spectrum has a thermal emission component, ii) the \chan\ quiescent spectrum is of sufficient quality to test if there is a hard emission tail, iii) the source has a reasonably accurate distance estimate (e.g. from X-ray bursts). 

Requirement i) listed above rules out some well-studied sources that have power-law dominated quiescent spectra, such as SAX J1808--3654 \citep[e.g.][]{Campana2002,Heinke2007,Heinke2009}, or NSs located near the Galactic center for which virtually all soft thermal emission is absorbed due to the very high interstellar extinction \citep[$N_{\mathrm{H}} \gtrsim 5\times 10^{22}~\nh$; e.g.][]{Degenaar2012_GC}. 
In practice, requirement ii) often implies that a source needs to be relatively bright or have multiple observations taken in quiescence. This rules out quite a number of NS LMXBs that were observed only once in quiescence by the \chan\ observatory, have low signal-to-noise spectra, or are not detected \citep[e.g.][]{Tomsick2004,Cornelisse2007,Jonker2007,Jonker2007_1H1905,Degenaar2017}. Finally, requirement iii) unfortunately rules out a source like MAXI J0556--332, which is bright and well-studied in quiescence, but for which the available distance estimates are not model independent and rely on some assumptions about the $M$ and $R$ of the NS in the system \citep[][]{Homan2014,Parikh2017_maxi}. 

Based on the above three criteria, we initially selected seven NS LMXBs for our analysis. Four of these, \ks, \xte\, \mxb\ and \exo, have been observed multiple times in quiescence to study the thermal evolution of the NS crust after an outburst \citep[see][for a review and Appendix~\ref{appendix:src} for source-specific references]{Wijnands2017}. 
In addition to these crust-cooling sources, the relatively nearby NS LMXBs \cen\ and \aql\ have been observed in quiescence many times \citep[e.g.][for compilations]{Cackett2010,Cackett2011,Campana2014} and are thus good targets for our study. Lastly, we included \uu\ in our analysis. Although this NS has been observed in quiescence only once, its quiescent spectrum is of decent quality and some constraints on its mass and radius have previously been obtained from the analysis of its X-ray bursts \citep[][]{Poutanen2014} and disk reflection \citep[][]{Degenaar2015}. It therefore seemed a promising target for our study. 
The seven sources in our sample are briefly described in Appendix~\ref{appendix:src}. The main properties that are relevant for the present work are summarized in Table~\ref{tab:srcprop}.

\section{Data reduction and analysis}\label{sec:reduction} 
Table \ref{Observations} gives an overview of the observations included in our study. We note that \ks, \xte\ and \mxb\ have more \chan\ observations in quiescence  than we used here. This is because for \ks\ and \xte\ we soon found that the small number of photons collected at energies $\lesssim$3~keV (due to a low flux and/or relatively high absorption column density) did not allow us to obtain meaningful constraints on the NS mass and radius (see Section~\ref{subsec:statistics}). In case of \mxb, the spectra taken at later times had significantly lower quality due to cooling of the NS crust \citep[and hence fading of the thermal X-ray emission;][]{Cackett2006,Cackett2013}. Therefore we decided not to include more spectra for those three sources. For \exo, \cen, \aql\ and \uu, we did use all available \chan\ data.

\begin{table*}
\centering
\begin{tabular}{ l  l  l  l l l l l}
\hline 
{Source name} & $\nu_\mathrm{spin}$ & $P_{\mathrm{orb}}$ & $D$ & Reference $D$ & $N_{\mathrm{H}}^{\mathrm{Gal}}$ & Previous $R$ (and $M$) constraints & Reference $R$ (and $M$)	\\
{} & (Hz) & (hr) & (kpc) &  & ($10^{21}$~cm$^{-2}$) &  & 	\\
\hline
\ks\ & 524 & - & $\lesssim$8 & [1,2] & 3.1 & $R<12$~km, $M<1.8~\Msun$ & [2] \\
\xte\ & - & - & $8.8\pm1.3$ & [3] & 7.2 & - & - \\
\mxb\ & 567 & 7.1 & 7--15 & [4,5] & 1.8 & - & - \\ 
\exo\ & 552 & 3.8 & $7.1\pm1.2$ & [5] & 1.0 & $M=2.00^{+0.07}_{-0.24}~\Msun$ and $R=11.3^{+1.3}_{-1.0}$~km & [6] \\
  &  &  &  &  &  &  (0.3--10 keV) & [6] \\
 &  &  &  &  &  & $M=1.50^{+0.4}_{-1.0}~\Msun$ and $R=12.2^{+0.8}_{-3.6}$~km & [6] \\
  &  &  &  &  &  &  (0.5--10 keV) & [6] \\
%
Cen X-4 & - & 15.1 & $1.2\pm 0.3$ & [7] & 0.9 & $R=7-12.5$~km for $M=1.5~\Msun$ & [8] \\
Aql X-1 & 550 & 19 & 3.5--8 & [5,9,10] & 2.8 & see Section~\ref{subsec:aqlx1} & [11] \\
\uu\ & 620 & $\sim$12? & $2.9-7.8$ & [5] & 18 & $R>12$~km for $M=1.0-2.4~\Msun$ & [12] \\
 &  &  &  &  &  & $R<21$~km for $M=1.5~\Msun$ & [13] \\
\hline
\end{tabular}
\caption{Summary of the main properties of the NS LMXBs in our sample relevant for the present study. 
For reference, we quote the Galactic absorption in the direction of each source, $N_{\mathrm{H}}^{\mathrm{Gal}}$, which reflects the weighted average value of \citet{Kalberla2005}.
References: 1=\citet{Muno2000_ks}, 2=\citet{Ozel2012}, 3=\citet{Lin2009}, 4=\citet{Oosterbroek2001}, 5=\citet{Galloway2008}, 6=\citet{Cheng2017}, 7=\citet{Chevalier1989}, 8=\citet{Cackett2010}, 9=\citet{Rutledge2001}, 10=\citet{Matasanchez2017}, 11=\citet{Li2017}, 12=\citet{Poutanen2014}, 13=\citet{Degenaar2015}.
}
\label{tab:srcprop}
\end{table*}

\begin{table*}
\centering
\begin{tabular}{ l  l  l  l  l}
\hline 
{Target}	& {ObsID} & {Date} & {Reference} & {Net counts}	\\
	&  & (mm-dd-yyyy) & & 	\\
\hline
{KS 1731--260}	&	{2428} & {03-27-2001} & {[1--7]} & {174$\pm$13}\\
{XTE J1701--462} & {7513} & {08-12-2017} & {[8,9]} & {200$\pm$14}\\
{MXB 1659-298} &	{2688} & {10-15-2001} & {[10--13]} & {30.35$\pm$5} \\
 &	{3794} & {10-16-2002} & {} & {259(16)} \\ 
{EXO 0748--676} &	{9070} & {10-12-2008} & {[14--16]} & {3190$\pm$50}\\ 
&	{10783} & {10-15-2008} & {} & {3150$\pm$60} \\ 
&	{9072}  & {06-10-2009} & {} & {4280$\pm$60} \\ 
&	{10871} & {02-25-2009} & {} & {1710$\pm$40}\\ 
&	{11059} & {04-20-2010} & & {3920$\pm$40}\\
& {12414} & {07-02-2011} & {} & {5430$\pm$70} \\ 
&	{11060} & {10-20-2013} & & {3760$\pm$60} \\
& {14663} & {08-01-2013} & & {3900$\pm$60} \\
{Cen X-4}	&	{713}	 & {06-23-2000} & {[17,18]} & {2670$\pm$50} \\
& {4576} & {06-21-2004} & {} & {960$\pm$30} \\ 
{Aql X-1} &	 {708} & {11-28-2000} & {[19--24]} & {1230$\pm$30}	\\ 
&{709}&{02-19-2001} & {} & {740$\pm$30} \\ 
&{710}&{03-23-2001} & & {860$\pm$30} \\
&{711}&{04-20-2001} & & {1140$\pm$30}\\
& {3484} & {05-05-2002} & {} & {1060$\pm$30}\\ 
&{3485}&{05-20-2002} & & {1280$\pm$40}\\
&{3486}&{06-11-2002} & & {2250$\pm$50}\\
& {3487} & {07-05-2002} & & {580$\pm$20}\\
& {3488} & {07-22-2002} & & {590$\pm$20}\\
&{3489}&{08-18-2002} & & {580$\pm$20} \\
&{3490}&{09-03-2002} & & {720$\pm$30}\\
&{12457}&{10-22-2010} & {} & {1620$\pm$40}\\ 
&{12458}&{10-30-2010} & & {920$\pm$30}\\
&{12459}&{11-12-2010} & & {1000$\pm$30}\\
{4U 1608-52}	&	{12470} & {10-11-2010} &  {this work} & {930$\pm$30}\\
\hline
\end{tabular}
\caption{List of all \chan\ ACIS-S observations used in this work. References: 1=\citet{Wijnands2001_ks}, 2=\citet{Wijnands2002}, 3=\citet{Rutledge2002KS}, 4=\citet{Cackett2006}, 5=\citet{Cackett2010KS}, 6=\citet{Zurita2010}, 7=\citet{Merritt2016}, 8=\citet{Fridriksson2010}, 9=\citet{Fridriksson2011}, 10=\citet{Wijnands2003}, 11=\citet{Wijnands2004}, 12=\citet{Cackett2008}, 13=\citet{Cackett2013}, 14=\citet{Degenaar2009}, 15=\citet{Degenaar2011}, 16=\citet{Degenaar2014}, 17=\citet{Cackett2010}, 18=\citet{Rutledge2001Cen},19=\citet{Rutledge2001}, 20=\citet{Rutledge2002}, 21=\citet{Campana2003}, 22=\citet{Li2017}, 23=\citet{Cackett2011}, 24=\citet{Campana2014}.
}
\label{Observations}
\end{table*}

All observations listed in Table~\ref{Observations} were obtained with the ACIS-S, sometimes using a sub-array to avoid pile-up. Exposure times varied between 4.7--42.9 ks for individual spectra. We refer the reader to the references in Table~\ref{Observations} for details on the different observations. We here only quote the full details of the single \chan\ observation of \uu, as to our knowledge those data have not been published previously. \chan\ observed \uu\ on 2010 October 11 starting at 03:15 \textsc{ut} for a duration of 23~ks. The observation was performed with the ACIS-S in a 1/8 sub-array and using the faint, timed data mode. There were no background flares during the observation, so all data could be used in the analysis. 

Data processing, reduction and spectral extraction of all observations was performed within \textsc{ciao} v. 4.9. The raw data were first put through a number of reprocessing steps consisting of applying the latest available calibration, removing episodes of background flaring and applying good time intervals. After those steps, the reprocessed data were suitable for spectral extraction. Source spectra were obtained using a $\sim 1''$-radius circular region, with the exact size depending on the source brightness. 
A background spectrum was extracted from the same image using a circular region with a radius three times larger than that of the source and placed well away from it. The routine \textsc{specextract} was used to create the spectra and build the ancillary response files (arfs) and the redistribution matrix files (rmfs). All spectra were rebinned using the tool \textsc{grppha} into channels with a minimum of 20 counts per bin, so that $\chi^2$ statistics could be used to probe the goodness of the spectral fits. 

Since \mxb\ displays X-ray eclipses we reduced the exposure time of the spectra by 900~s per eclipse, to make sure that we obtain the correct X-ray flux (and spectral parameters) during our fits. Similarly, the spectra of \exo\ were corrected; in this source the eclipses last for 500~s. To determine the times of eclipses during our observations we used the ephemeris of \cite{Oosterbroek2001} for \mxb\ and of \cite{Wolff2009} for \exo.\footnote{In case of \mxb, the only \chan\ observation used in this work contained a single eclipse, so the exposure time was reduced with 900~s accordingly. In case of \exo\ we reduced the exposure time with 500 s for observations 9070, 10783, 10871, with 1000 s for observations 9072, 11059, 11060 and with 1500 s for observations 12414 and 14663, depending on whether one, two or three eclipses occurred. }

We estimated the level of pile-up in each of the analyzed observations, using the web version of \textsc{pimms} (Portable
Interactive Multi-Mission Simulator), which uses the proportionality between
the pile-up fraction and the true count-rate to estimate the level of pile-up in a data set. For none of the data considered in this work the pile-up fraction exceeded 6$\%$, which we deemed sufficiently small to leave uncorrected in the scope of the present work. Although it was demonstrated that even a pile-up level of 1$\%$ can affect the $M$ and $R$ constraints \citep{Bogdanov2016}, we are not after precision measurements in this work and the effect of pile-up will be much smaller than other biases such as e.g. the distance uncertainty. 
 
All spectral fits were performed within \textsc{XSpec} (v. 12.9.1), ignoring all data outside the energy range of 0.3--10 keV. Five of the sources in our sample (\mxb, \exo, \cen, \aql\ and \uu) have orbital information that rule out an ultra-compact nature (see Appendix~\ref{appendix:src}). For the remaining two sources (\ks\ and \xte) there are no direct constraints on their orbital parameters, although their X-ray burst or accretion properties suggest that these both accrete H-rich material too \citep[][]{Galloway2008,Lin2009}. Moreover the luminous outbursts experienced by both sources seem to disfavor very small orbits and hence H-deficient donor stars \citep{Shahbaz1998,Meyer2004,Wu2010}.\footnote{In any case, we could not obtain constraints on $M$ and $R$ from the quiescent thermal emission of \ks\ and \xte. The uncertainty in their atmosphere composition is thus not a concern for the present work.} A H-atmosphere should thus be appropriate to model the thermal spectra of the NSs in our sample. We choose to use \textsc{nsatmos} \citep[][]{Heinke2006}, which has five fit parameters: the NS effective temperature ($T_{\mathrm{eff}}$), mass ($M$), and radius ($R$), which are all in the NS frame, as well as the source distance ($D$) and a normalization factor. The latter parametrizes the fraction of the surface that is emitting the thermal radiation; this parameter was fixed to 1 in all fits (i.e. we assume a homogeneously emitting NS). In addition to the thermal emission component, we tested for and modeled any possible hard emission tail using \textsc{pegpwrlw}. This model is characterized by a power-law index $\Gamma$ and a normalization that gives the unabsorbed flux of the power-law component in the energy range that is specified by the remaining two parameters (for which we took 0.5--10 keV). Finally, we modeled the interstellar absorption ($N_{\mathrm{H}}$) using the \textsc{tbabs} model \citep{Wilms2000} with \textsc{vern} cross-sections \citep{Verner1996} and \textsc{angr} abundances \citep{Anders1989}.  

To reduce the number of free parameters during the spectral fits, the distance to the source was always fixed (but we did probe a distance range) to a value compatible with that reported in literature (see Table~\ref{tab:srcprop}). This strategy leaves $N_{\mathrm{H}}$, $T_{\mathrm{eff}}$, $M$ and $R$ as free parameters, together with $\Gamma$ and the power-law normalization in case a hard emission tail was modeled. When multiple spectra were available for a single source, those spectra were fitted simultaneously with the parameters that are not expected to change over time (e.g. the $N_{\mathrm{H}}$, the distance, NS mass and radius) tied between the different data sets. To reduce the number of degeneracies, we also decided to tie the $\Gamma$ indices relative to the various spectra. However, the NS temperature and power-law normalization were always allowed to freely vary between different data sets, as those are known to change over time for all our sources. 

For each source, our first step was to choose the right model to describe the spectrum and to evaluate the need for including a power-law component in the spectral fits. To this aim, we fitted each of the observations with both a single thermal model, \textsc{tbabs}$\times$\textsc{nsatmos}, and one with an additional hard power-law tail,  \textsc{tbabs}$\times$(\textsc{nsatmos+pegpwrlw}). We then employed the standard practice to test the significance of the non-thermal component via an f-test (\textsc{ftest} in \textsc{XSpec}), which estimates the probability that the improvement in $\chi^2$ due to the addition of  an extra component, in our case a power-law, to the model is occurring by chance. If the obtained f-test probability was on the order of 10$^{-3}$ or larger, we considered the contribution of the non-thermal component not significant. The power-law component was kept in the model when the f-test results gave a much smaller probability. 

Without very good statistics, the attempt of constraining the NS mass and radius at the same time  usually resulted in huge error bars on all spectral parameters. We therefore applied an alternative approach in which we focused on obtaining the radius of the NS and probed how the best-fit radius changed when stepping the mass through a range of reasonable values. In other words, in this strategy several fits were launched for the same spectrum leaving, in each of them, $M$ fixed to a different value in the range of $1-2~\Msun$. This range is motivated based on the approximate range between the lowest measured NS mass \citep[one of the NSs in the double-NS system J0453+1559, with a mass of $1.174\pm0.004~\Msun$;][]{Martinez2015} and the highest one \citep[the current record-holder J0348+0432 with a mass of $2.01\pm0.04~\Msun$;][]{Antoniadis2013}.  

All the best-fit radii individuated by the fits form a wide range which, under the hypothesis that the model correctly describes these spectra,  includes the real radius of the NS. The associated errors to the estimated values were initially calculated using the \textsc{error} command, which individuates, for a specified parameter, the  90$\%$ confidence region of it, exploring the $\chi^2$ landscape around the best-fit radius. We then improved the precision of these results applying a Goodman-Weare algorithm of Monte Carlo Markov Chain \citep[MCMC;][]{Goodman2010}, appropriate in the case of a skewed and elongated distribution for mass and radius, using the command \textsc{chain} in \textsc{XSpec}. Whenever the mass value was frozen, the MCMC procedure was launched with 300 walkers and a chain length of 9000, whereas when both $M$ and $R$ were left free in the fit, we used 200 chains, each of length equal to 2$\times$10$^6$. 
The errors reported in this work, unless otherwise specified, were calculated with the \textsc{error} command for the resulting simulated posterior distribution and reflect 90\% confidence intervals.

\section{Chandra results}\label{sec:results}
Out of the seven sources in our sample, we were able to obtain meaningful constraints on $R$ (and $M$) for only two NS LMXBs, \exo\ and \aql, which we describe in Section~\ref{subsec:MRconstraints}. For the other five sources it was not possible to constrain the NS parameters. In Section~\ref{subsec:statistics} we argue that the main limitation is the number of counts in the spectra.  
Nevertheless, the search for exact NS masses and radii was not the main aim of this work; rather we wanted to test how constraining and reliable the results for field LMXBs may be in general, and in particular to probe the effect of some of the biases mentioned in Section~\ref{sec:intro}. The data quality for \exo\ and \aql\ allowed us to explicitly test the effect of the energy range over which the spectra are fitted, the presence of a power-law spectral component, and distance uncertainties. We present and discuss these results in Sections~\ref{subsec:Erange}, \ref{subsec:aql} and \ref{subsec:distance}, respectively.

\subsection{The effect of limited statistics}\label{subsec:statistics}
The failure to obtain strong constraints on the stellar parameters for most of the NSs in our sample likely has different reasons. To this extent, we can divide our targets into different groups. The first group is composed of \ks, \xte\ and \mxb; these three NSs have been observed with \chan\ multiple times in quiescence and hence appeared good targets for our study. However, as can be seen in Table~\ref{Observations}, a limited number of counts ($\lesssim 300$ counts) are obtained for these three sources due to their low fluxes. In case of  \xte\ the absorption column is also relatively high ($\sim 10^{22}~\nh$; see Table~\ref{tab:srcprop}), which further reduces the thermal flux and hence any constraints on the NS parameters obtained from it. Furthermore, \xte\ contains a significant power-law component, which further adds to the uncertainties (see below). Nevertheless, \ks\ and \mxb\ have a lower column density and purely thermal spectrum, but still no constraints on $M$ and $R$ can be obtained. Clearly, having only a few hundreds of counts per spectrum is not sufficient to obtain any meaningful constraints on the NS parameters. 

The second group of sources is then composed of \cen\ and \uu. The spectra obtained for these sources are of much higher quality ($\sim 900-2700$ counts per spectrum) than that of the first group, yet no strong constraints on $M$ and $R$ can be obtained. For \uu\ we suspect that its relatively high absorption ($\sim 10^{22}~\nh$) plays a role, but this is certainly not the case for Cen X-4 (see Table~\ref{tab:srcprop}). What might be an issue is that these two sources contain a significant power-law component in their spectra; this adds two free parameters in the spectral fits (the power-law index and normalization) and may also create degeneracies  (e.g. a positive correlation between radius and power-law index) that limit the constraints that can be obtained on the NS parameters. 

In case of Cen X-4, the power-law spectral component contributes $\sim$50\% to the total unabsorbed 0.5--10 keV flux \citep[see also][]{Cackett2010}, whereas this is $\sim$20\% for \uu. Nevertheless, Aql X-1 also contains a similarly strong power-law component in some of its X-ray spectra \citep[e.g.][see also Section~\ref{subsec:aql}]{Cackett2011} and yet we do obtain good constraints for that NS. Since Aql X-1 has been observed with \chan\ in quiescence much more often (14 times) than \uu\ (only once) and Cen X-4 (only twice), this suggests that statistics is  the main limiting factor. This is confirmed by our \athena\ simulations presented in Section~\ref{sec:simulations}, where we found that strong constraints on the NS parameters can be obtained for \uu\ with much better data, despite its relatively high column density and considerable power-law spectral component. It thus appears that particularly in case of a power-law emission tail and a high absorption column density, spectra with at least several thousands of counts are required to measure the NS parameters.

Having put up front that no constraints can be obtained for any of the NSs in these five quiescent LMXBs, we briefly give some more detail about the results obtained for the individual sources. For \ks\ and \mxb, the high value of the f-test probability suggested their spectra could be described by a purely thermal model. \xte, \cen\ and \uu, on the other hand, were modeled with the inclusion of a power-law component. For \ks, \xte\ and \uu, the errors of $R$ pegged at the lower and upper limits hence no probable radius range could be isolated. 

For \mxb\ we were able to obtain limits on the NS radius, but only with very large errors; the best-fit radii are comprised in the range 6--20~km for NS masses of 1--2~M$_\odot$. For \cen, our analysis procedure led to a range of 5--17~km for the radius of its NS (for $M=$1--2~M$_\odot$). We can compare these results to that of \citet{Cackett2010} where the radius of the \cen\ was measured to be $R=7-12.5$~km for $M=1.5~\Msun$ at a confidence level of 68$\%$, via the analysis of combined \chan, \xmm, \suzaku and \asca observations. This is compatible with our result; assuming a mass of $M=$1.5~M$_\odot$ for Cen X-4, we obtain a radius measurement of $R=5-15$~km at a confidence level of 90$\%$. For both \mxb\ and Cen X-4 the obtained radius ranges are so wide that they comprise basically all EoSs ever hypothesized, so no interesting constraints are obtained. 

\subsection{Constraints for \exo\ and Aql X-1}\label{subsec:MRconstraints}
\begin{figure}
\centering
\includegraphics[height=6 cm, width=8 cm]{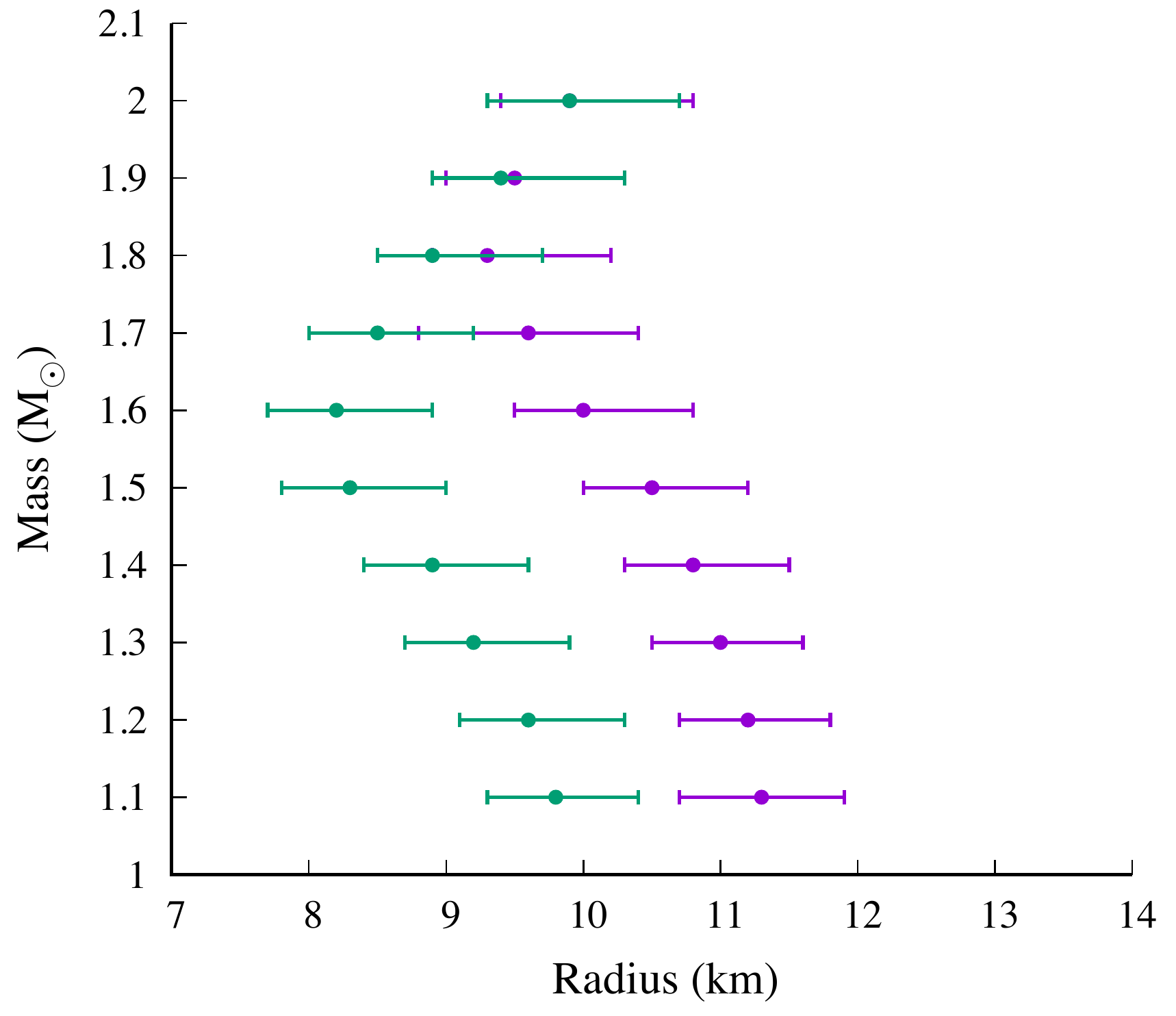}
\includegraphics[height=6 cm, width=8 cm]{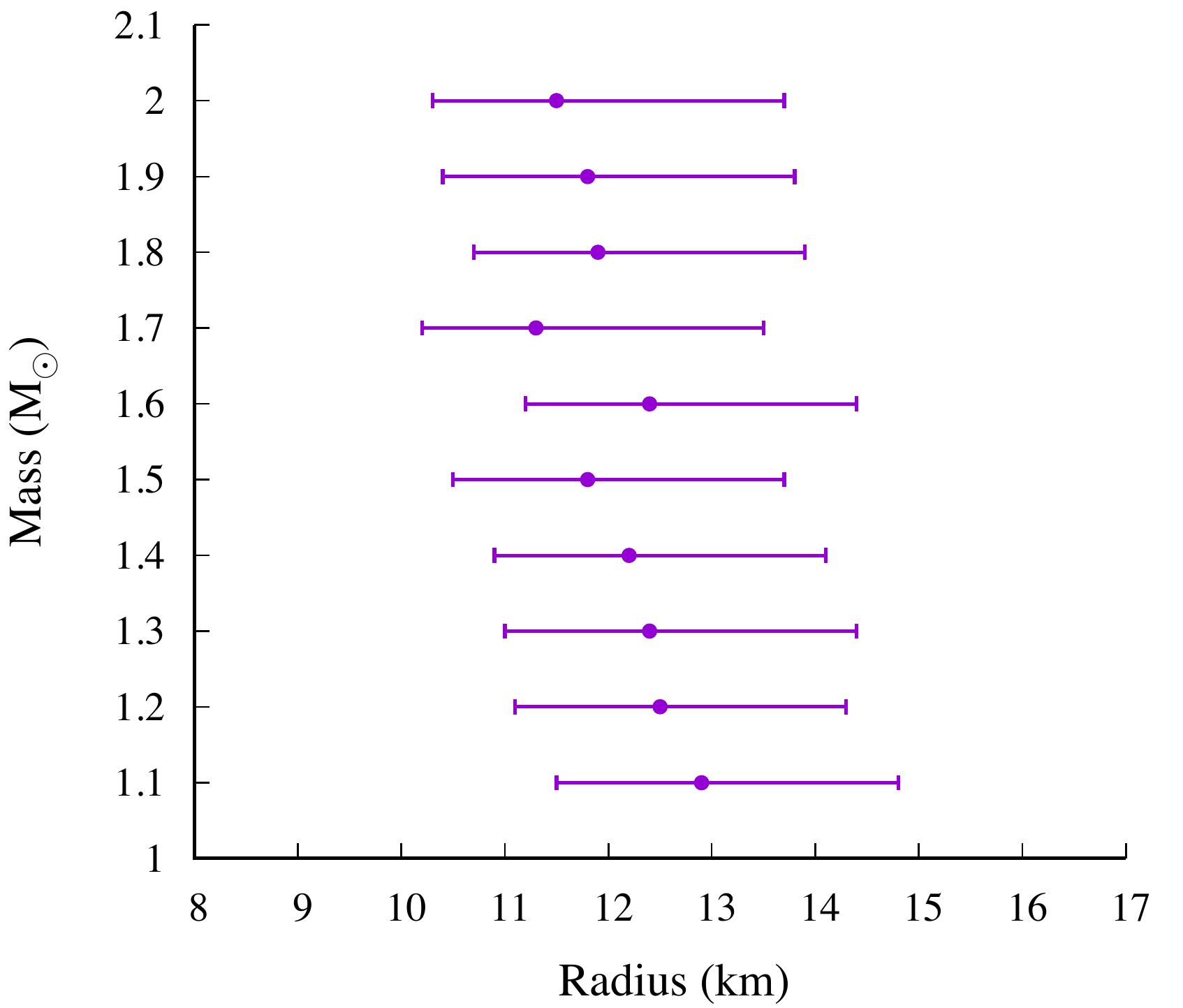}
\caption{Radius constraints for \exo\ (top) and \aql\ (bottom) for different mass values. As detailed in Section~\ref{subsec:Erange}, for \exo\ we investigated fits performed over two different energy ranges of 0.3--10 keV (in purple) and 0.5--10 keV (in green). For all other analysis, including that of Aql X-1 shown here, only an energy range of 0.3--10 keV was considered. Error bars reflect 90\% confidence levels.
}
\label{fig:EXOAQLresults}
\end{figure}

For all eight spectra of \exo, an f-test does not yield evidence of a strong power-law component in the spectrum. We therefore fitted all data of this source using a NS atmosphere model only. Due to the high number of counts in the spectra (see Table~\ref{Observations}), the final results for the radius are comprised in a relatively small range of 9--12~km (see Table~\ref{tab:REXO}), which can be seen in Figure~\ref{fig:EXOAQLresults} (top, purple). The full results for these fits can be found in Table~\ref{tab:REXO}. The quality of the data allowed us to  find constraints on both the mass and the radius of the NS, yielding $M=$1.65$^{+0.11}_{-0.80}$~M$_\odot$ and $R=$10$^{+2}_{-1}$~km for an assumed distance of 7.1~kpc (see Table~\ref{tab:MRFinal}). These results are consistent within the (large) errors with the findings of \citet{Cheng2017} using \xmm\ quiescent data (see Section~\ref{subsec:Erange}). For \exo\ we investigated the effect of the energy range considered for the spectral fits on the obtained NS mass and radius. These results are included in Figure~\ref{fig:EXOAQLresults} (top, green) and Table~\ref{tab:MRFinal}, but are discussed in detail separately in Section~\ref{subsec:Erange}. In Section~\ref{subsec:distance} we discuss the effect of the distance uncertainty on the radius constraints for \exo.
 
Using an f-test as a diagnostic of the need to include a power-law spectral component, we found that for Aql X-1 the data could be split into two samples, depending on whether or not they required a power-law component (see Section~\ref{subsec:aql}). However, as we argue in Sections~\ref{subsec:aql} and~\ref{subsec:distance}, it appears that likely all spectra of Aql X-1 include a hard emission tail. Therefore, we here report on the mass and radius results obtained for this NS by fitting all of its 14 \chan\ spectra with a composite model of a NS atmosphere and an additional power-law component. The results of these fits are shown in Figure~\ref{fig:EXOAQLresults} (bottom), and yield radii between 10 and 15 km (see also Table~\ref{RAqlX1All}). Leaving the mass to be a free fit parameter and assuming a distance of 5~kpc (see Section~\ref{subsec:distance} for a discussion on the distance bias), we were able to constrain the mass and radius simultaneously, albeit with large errors: $M=$1.8$^{+0.4}_{-0.9}~\Msun$ and $R=$10.2$^{+3.2}_{-1.5}$~km (see Table~\ref{tab:MRFinal}).

\subsection{EXO 0748--676: Dependence on the fitted energy range}\label{subsec:Erange}
In all our analysis we fitted the \chan\ spectral data over an energy range of 0.3--10~keV. Recently, \citet{Cheng2017} used \xmm\ data of \exo\ in quiescence to infer the NS parameters, and found that the results strongly depended on the energy range over which the spectral fits were performed. Fitting the \xmm\ data to a NS atmosphere model, the authors obtained $M=2.00^{+0.07}_{-0.24}~\Msun$ when fitting over 0.3--10 keV and $R=11.3^{+1.3}_{-1.0}$~km, and $M=1.5^{+0.4}_{-1.0}~\Msun$ and $R=12.2^{+0.8}_{-3.6}$~km when using a range of 0.5--10 keV (at a confidence level of 90$\%$). 

Prompted by these results, we tested if the same effect is seen for \chan\ data to rule out that it is instrument-specific. Since we performed all our fits initially in the energy range of 0.3--10~keV (see Section~\ref{sec:reduction}), we repeated our fits for \exo\ but now ignoring all data in 0.3--0.5 keV range. 

We found that the 0.5--10 keV fits performed for fixed masses resulted in radii of 8--10.5~km for \exo, slightly lower than the results of 9--12~km found for spectra fitted in the original energy range (0.3--10~keV). This can be seen in Figure~\ref{fig:EXOAQLresults} (top), whereas the full details of these fit results are included in Table~\ref{tab:REXO}. As can be seen in Table~\ref{tab:MRFinal}, the fit results with the mass and radius both free, also yielded different values than when considering the 0.3--10 keV range, albeit still compatible within the large error bars. We thus also find a lower mass and radius when fitting over 0.5--10~keV rather than over 0.3--10~keV, similar to what \citet{Cheng2017} found when analysing \xmm\ data. 

We note that when comparing the fits over the different energy ranges, deviations are also seen in the best-fit values for $N_{\mathrm{H}}$ and $kT_{\mathrm{eff}}$. In Tables \ref{tab:MRFinal} and \ref{tab:REXO} it can be seen that where the outcomes of the fits diverge the most, an analogous difference appears in the $N_{\mathrm{H}}$ and $kT_{\mathrm{eff}}$ values, being lower for the 0.5--10 keV results than for the 0.3--10 keV ones. 

However, as we show in Section~\ref{sec:simulations}, \athena\ simulations for \exo\ result in exactly the same mass and radius pairs for the 0.3--10 and 0.5--10 keV energy range. This suggests that statistics plays an important role in the obtained discrepancy. We note that Aql X-1 was not suitable to perform the same test because of the limited number of counts below $\simeq$0.5~keV in the \chan\ data we considered here (presumably due to its factor $\sim 3$ higher hydrogen column density than in \exo; see Table~\ref{tab:srcprop}).

 \begin{table*}
\centering
\begin{tabular}{l l l l l l l}
\hline 
{Source (energy range)}&{$M$} & {$R$}&{$\delta_{\mathrm{dist}}$} & {$N_{\mathrm{H}}$} & {$\log T_{\mathrm{eff}}$} & {$\chi^2_\nu$ (d.o.f.)} \\
 &($M_{\odot}$) & (km) & (km)& (10$^{21}$ cm$^{-2}$) &  \\
\hline
{EXO 0748--676 (0.3--10~keV)} & {1.65$^{+0.11}_{-0.80}$}&{10.0$^{+2.0}_{-1.0}$}& $\sim$ 2.5 &{0.51$^{+0.07}_{-0.09}$} & {6.30$^{+0.04}_{-0.08}$} &{1.14(653)}\\
{EXO 0748--676 (0.5--10~keV)} & {1.40$^{+0.13}_{-0.80}$}&{8.9$^{+1.7}_{-1.0}$}& $\sim$ 2.5 & {0.21$\pm$0.11} & 6.35$^{+0.01}_{-0.12}$ &{1.07(615)}\\
{Aql X-1 (0.3--10 keV)}&{1.80$^{+0.40}_{-0.90}$}&{10.2$^{+3.2}_{-1.5}$}& $\sim$ 6 &{4.10$\pm$0.40}& {6.29$^{+0.04}_{-0.10}$}& {0.87(540)} \\
\hline
\end{tabular}
\caption{Simultaneous constraints on the mass and radius of the NSs in \exo\ and \aql, assuming distances of 7.1 and 5~kpc, respectively. Quoted errors reflect 90\% confidence levels. $\delta_{\mathrm{dist}}$ gives an estimate of the additional error in the radius when the distance uncertainty is taken into account.}
\label{tab:MRFinal}
\end{table*}

 \begin{figure}
\centering
\includegraphics[height=6 cm, width=8 cm]{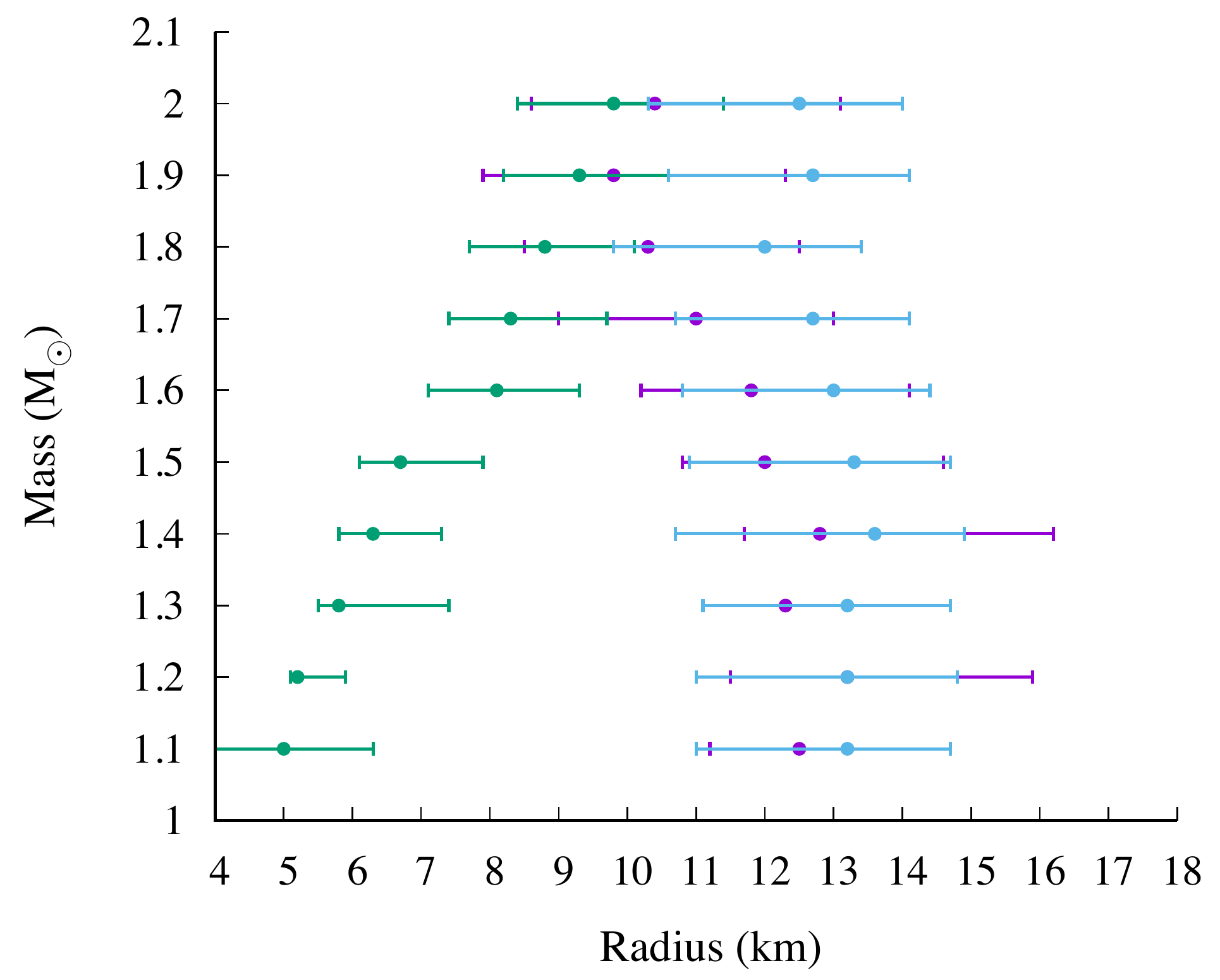}
\caption{Range of obtained radii for the NS in Aql X-1 for different NS masses. In purple the "Power-Law" sample, in green the "No Power-Law" sample, and in blue the results for the spectra originally assigned to the "No Power-Law" sample after adding a power-law component. In all the performed fits the distance value was kept fixed at 5 kpc. Error bars reflect 90\% confidence levels.}
\label{fig:AqlX1}
\end{figure}

\subsection{Aql X-1: Dependence of a power-law component}\label{subsec:aql}
Depending on whether a relevant fraction of the quiescent emission cannot be ascribed to a soft thermal component, we have two different models for describing the observations: a pure NS atmosphere model or a NS atmosphere model with an additional power-law tail. It is interesting then to ask ourselves if both types of spectra are equally suitable for obtaining mass and radius measurements of the NS. Aql X-1 is ideally suited to carry out this comparison; it has been observed multiple times in quiescence (see Table~\ref{Observations}) and has a variable spectral shape that is sometimes mostly thermal and at other times contains a significant power-law component that may contribute up to $\sim$80\% of the total unabsorbed 0.5--10~keV flux \citep[e.g.][]{Cackett2011}. 
Based on a simple f-test (see Section~\ref{sec:reduction}), every spectrum of Aql X-1 was assigned either to the "No Power-Law" (6 spectra) or the "Power-Law" (8 spectra) sample. The results are summarized in Table~\ref{tab:ftest}. If the NS is always uniformly emitting, we would not expect to find differences in the radii obtained for the two different samples. 

\begin{table}
\centering
\begin{tabular}{ l  l  l  }
\hline \hline
{ObsID} & {$f$-test ptobability} & {Assigned Sample}	\\
\hline
{708} & {0.10} & {No Power-Law}	\\
{709}&{8.8$\times$10$^{-5}$} & {Power-Law}\\
{710}&{5.2$\times$10$^{-6}$}& {Power-Law}\\
{711}&{3.4$\times$10$^{-9}$}&{Power-Law}\\
{3484} & {0.013} & {No Power-Law} \\
{3485}&{4.9$\times$10$^{-9}$}&{Power-Law}\\
{3486}&{6.7$\times$10$^{-17}$}&{Power-Law}\\
{3487} &{0.11}&{No Power-Law}\\
{3488} &{0.056}&{No Power-Law}\\
{3489}&{1.6$\times$10$^{-4}$}&{Power-Law}\\
{3490}&{2.4$\times$10$^{-5}$}&{Power-Law}\\
{12457}&{6.4$\times$10$^{-5}$}&{Power-Law}\\
{12458}&{1.9$\times$10$^{-3}$}&{No Power-Law}\\
{12459}&{1$\times$10$^{-3}$}&{No Power-Law}\\
\hline
\end{tabular}
\caption{Resulting f-tests probability for all observations of Aql X-1 and the resulting sample assignment. }
\label{tab:ftest}
\end{table}

For the "No Power-Law" sample we find a range of best-fit radii of 5--11.5 km for varying masses, as can be seen in Figure \ref{fig:AqlX1} (green). We note that the inferred radius increases with mass and that trends are also observed in the values of $N_{\mathrm{H}}$ and $kT_{\mathrm{eff}}$ (increasing and decreasing, respectively, for increasing mass values; see Table \ref{RAqlX1NPL}). These trends are not observed for the other Aql X-1 fits, where the $N_{\mathrm{H}}$ values tend to the similar values regardless of the chosen mass and the average temperature is positively correlated with $M$ (see Tables \ref{RAqlX1All}, \ref{RAqlX1PL}, and \ref{RAqlX1NPL12}).
Furthermore, although fits to the individual spectra do not suggest the presence of a significant power-law component in any of them, fitting the whole sample together gives a rather poor fit (see Table~{\ref{RAqlX1NPL}}) and suggest that an \textsc{nsatmos}-only model might not be the correct description of these data (see below). 
We carried out a similar analysis for the "Power-Law" sample of Aql X-1, which is presented in Figure~\ref{fig:AqlX1} as well (purple). We find a wide range of radii from $\approx$ 8 to 16 km. Notably, this range is overlapping with that of the "No Power-Law" sample only in a small part of this total range. There is thus a mismatch in the results obtained for the two samples. Furthermore, we do not find the same proportionality of $R$ with $M$ that we saw for the sample that we fitted with a NS atmosphere model only, nor the trends exhibited by the $kT_{\mathrm{eff}}$ and $N_{\mathrm{H}}$ parameters apparent in the "No Power-Law" sample fits. We note that we obtain a power-law index of $\Gamma \approx 1.1-1.3$. These fits are presented in Tables~{\ref{RAqlX1PL}}.

We considered whether an unmodeled power-law component could be the cause of the different results obtained for the two samples, and the bad overall fit for the "No Power-Law" sample. To test this, we repeated the analysis of that sample by fitting the data with a power-law component included. Due to the low number of counts at higher energies ($\gtrsim$3~keV) in these spectra, we had to fix the power-law component in these fits. We choose to set $\Gamma = 1.2$, which is the value that we obtained for the "Power-Law" sample (see Table~{\ref{RAqlX1NPL12}}) and is consistent with previous studies \citep[e.g.][]{Cackett2011,Campana2014}. As shown in Figure \ref{fig:AqlX1} (blue) the best-fit radii for these fits replicate results obtained for the "Power-Law" sample and the overall fit statistic is significantly improved. 

As discussed in Section~\ref{subsec:distance}, we found another peculiar effect for the "No Power Law" sample related to the (lack of a) distance dependence. Taken together, this leads us to believe that in spectra of the sample that were originally earmarked as not containing a significant power-law component, there is in fact an unmodeled hard emission tail that causes unexpected behavior in the fit results. This would suggest that finding a proportionality between the inferred radius and assumed mass can also serve as a test to spot an unmodeled power-law component in the spectra. 

Based on our findings above, we performed a final set of fits using all the 14 \chan\ observations of Aql X-1 with a composite thermal and power-law model, to constrain the mass and radius of this NS. The results of these fits were discussed in Section~\ref{subsec:MRconstraints} and are listed in Table~\ref{RAqlX1All}. We note here that the range of inferred radii for the final analysis of all 14 observations, comprised between 10 and 15 km, is only slightly smaller than the initial results from the "Power-Law" sample. Notably, the improvement of simultaneous constraints on mass and radius are not significant, despite that the number of spectra nearly doubled. This  is likely due to the fact that the addition of a power-law component increases the number of free fit parameters and the level of degeneracy. Nevertheless, the presence of a hard emission tail in itself does not necessarily limit the constraints that can be obtained for the NS parameters. As we show in Section~\ref{sec:simulations}, very high-quality spectra such as can be obtained with \athena, have the potential to yield accurate mass and radius measurements even when the quiescent spectrum contains a significant hard emission tail.

\begin{figure}
\flushleft
\includegraphics[height=6 cm, width=8 cm]{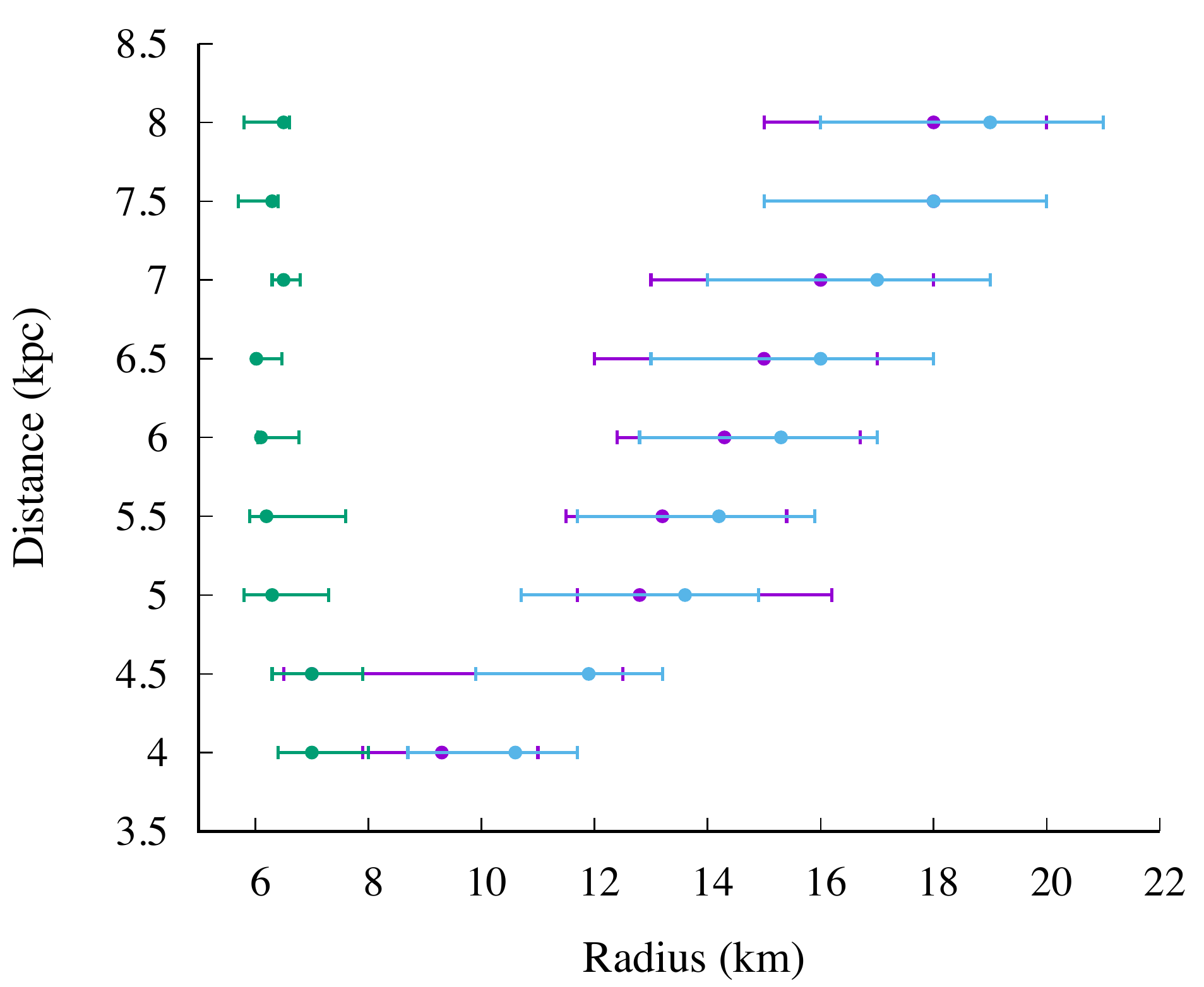}
\caption{Best-fit radii for Aql X-1 for different values of the distance. In purple, the "Power-law'' sample, in green the "No Power-law'' sample, and in blue the  sample indicated originally indicated as "No Power-law'' but now fitted with a power-law spectral component added (with a fixed index of $\Gamma=1.2$). In all the performed the fits the mass was kept fixed at 1.4 $\Msun$. Error bars reflect 90\% confidence levels.}
\label{fig:Distance}
\end{figure}

\subsection{The influence of the distance uncertainty}\label{subsec:distance}
The uncertainty about the distance of a source is one of the main issues of this technique. If the distance is not known accurately, a range of distance values can be chosen for the analysis and this will bias the results. Since the observed thermal flux is proportional to $(R/D)^2$, a larger distance should return a larger NS radius. To illustrate the magnitude of this bias we launched multiple fits for \exo\ and \aql, in which the distances were fixed to different values within the allowed range for each source. These ranges are 5.9--8.3~kpc for EXO 0748--676 (Section~\ref{subsec:exo}) and 4--8~kpc for Aql X-1 (Section~\ref{subsec:aqlx1}). In all these fits we fixed $M=1.4~\Msun$ to reduce the number of degrees of freedom. 

The expected proportionality relation between the NS radius and distance is indeed recovered for \exo. The best-fit radius goes from $\sim$8.5--9.5~km at a distance of 6.1 kpc to $R\sim$12--13~km at the other end of the distance confidence range (see Table~\ref{tab:REXODist}). This shows that allowing the full distance range gives an additional uncertainty of $\delta_{\mathrm{dist}}$$\sim$2.5~km for the NS radius.For Aql X-1, we performed fits with different distances for both the ``Power-law'' and ``No Power-law'' sample (see Section~\ref{subsec:aql} and Table~\ref{tab:ftest}). For the former we observe the same expected behavior as for \exo, where the inferred radius increases with assumed distance (see Figure~\ref{fig:Distance}, purple). We find radii between 8--11 km when using 4 kpc and 16--20~km when using 8~kpc (see Table~\ref{RAqlX1NPL}). Accounting for the distance uncertainty thus results in an additional error $\delta_{\mathrm{dist}}$$\sim$6~km on the best-fit radius.  
 
For the Aql X-1 spectra that we identified as not requiring a power-law spectral component, however, we do not find the expected proportionality between $R$ and $D$. In Figure~\ref{fig:Distance} (green) it can be seen that for this selection of spectra the obtained radius does not depend on the assumed distance.  In addition, the obtained radii are unphysically small for a NS. Failure to recover the expected correlation may indicate that the data are not modeled correctly, as is also suggested by the poor $\chi^2$ fit values obtained when the entire sample is modeled simultaneously (see Table~\ref{RAqlX1NPL}). To test this, we repeated the spectral fits of the original ``No Power Law'' after including a power-law component with a fixed index of $\Gamma = 1.2$ (see Section~\ref{subsec:aql}). Indeed, the proportionality between $R$ and $D$ is then seen (Figure~\ref{fig:Distance} blue; see also Table~\ref{RAqlX1NPL}), and is similar to the results obtained for the original ``Power Law'' sample. This seems to suggest again that despite the low f-test probability for the individual spectra, there might be a power-law tail contained in all these spectra after all. 

To conclude this part of our analysis, we performed one last series of fits to probe the distance bias for the all-inclusive sample, i.e. comprising all 14 \chan\ observations, fitted to a composite model that includes a power-law component. With radius estimates fluctuating from 9--12 to 16--20~km from one end to the other of the distance range, it is clear that the expected $D-R$ proportionality is again maintained (see Table~\ref{RAqlX1All}). 
Our results on Aql X-1 indicate that failure to recover the expected $D$ dependency may also serve as a test to see if there is an unmodeled power-law component in the data.

\begin{figure*}
\centering
\includegraphics[height=6 cm, width=8.5 cm]{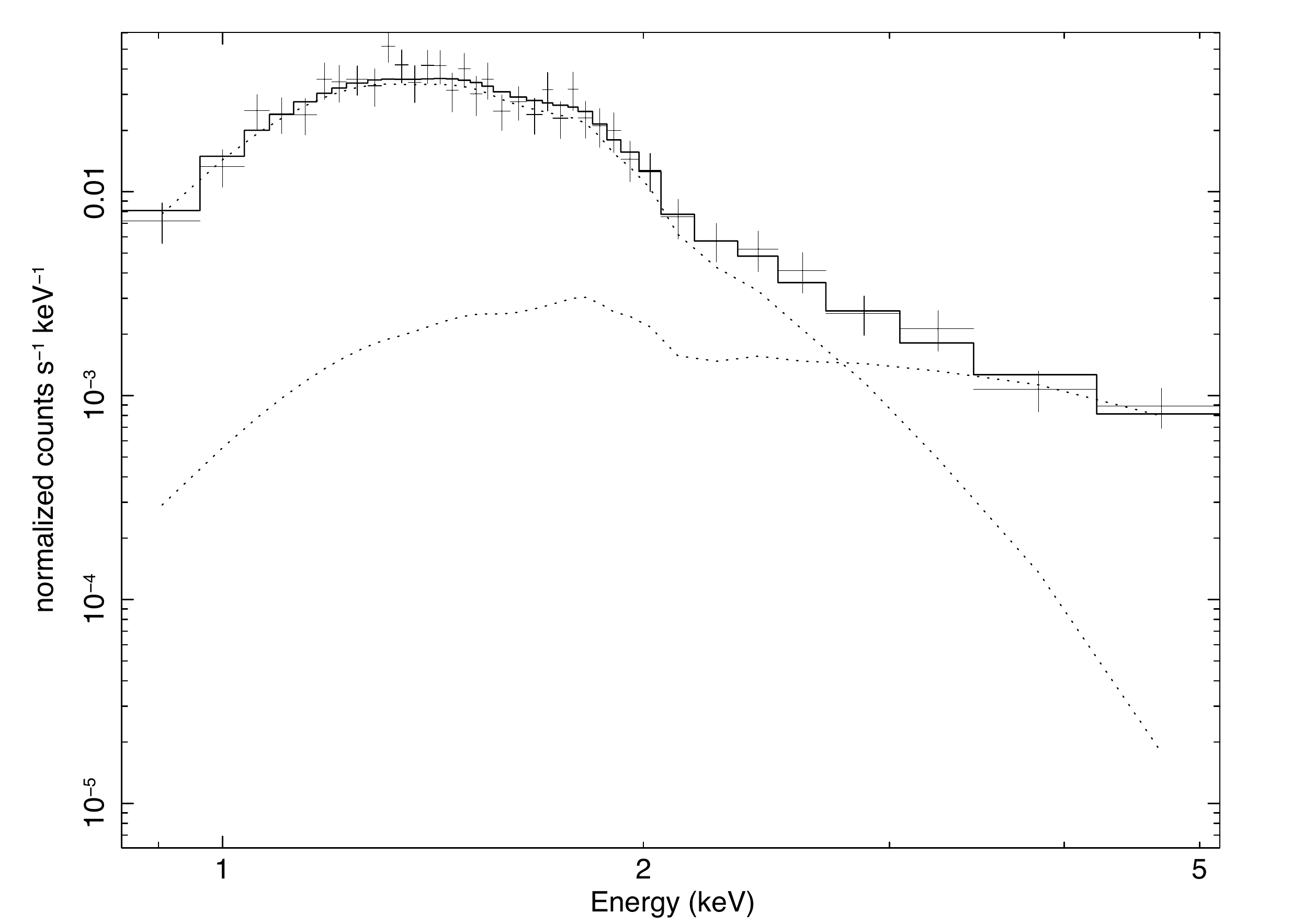}
\includegraphics[height=6 cm, width=8.5 cm]{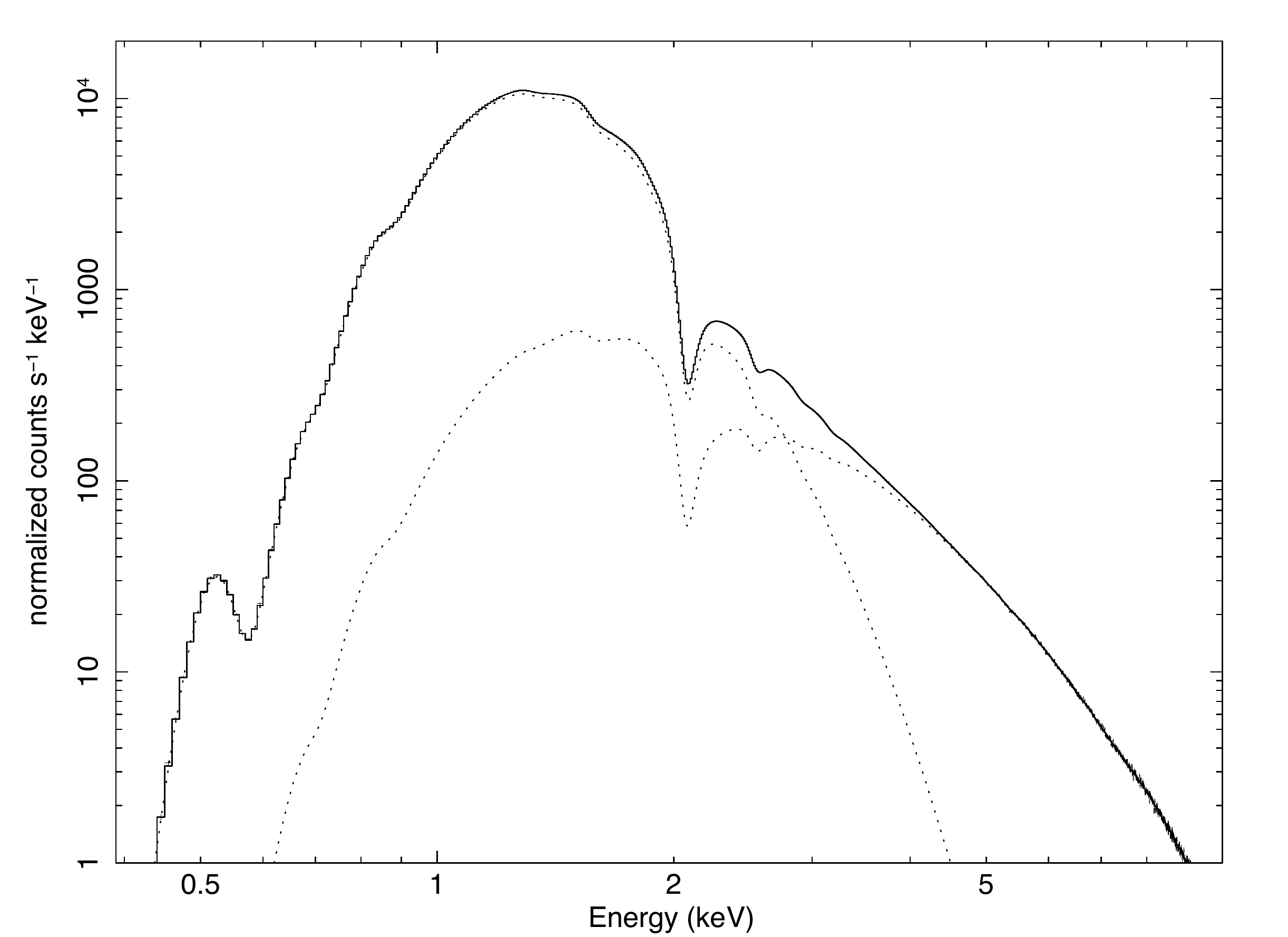}
\caption{Quiescent X-ray spectra of \uu. Left: Observed \chan/ACIS-S spectrum (exposure time of 23~ks). Right: Simulated \athena/WFI spectrum based on the model used to fit the \chan\ data (exposure time of 50~ks). The solid curves indicate fits to a composite model comprised of a NS atmosphere (dominating the emission at energies below $\sim$3~keV) and a hard power-law tail, which are indicated by the dotted curves.
}
\label{fig:Athena}
\end{figure*}

\section{Athena simulations}\label{sec:simulations}
Not surprisingly, our analysis shows that good statistics is key to obtain NS mass and radius constraints for field qLMXBs. It is then interesting to explore if such constraints can be improved with new instrumentation and if this can also help to lift certain biases. To this end we performed simulations for ESA's \athena mission, which is currently scheduled for launch in the late 2020s \citep[][]{Athena2015}. Its large collective area at soft X-ray energies is expected to be very useful for obtaining NS mass and radius constraints from studying their quiescent thermal X-ray emission.  Performing these simulations thus allows us to probe if and how much the limits of this technique depend on the quality of our spectra. 

All simulations were performed using the \textsc{Xspec} command \textsc{fakeit}, which builds a synthetic spectrum based on a given model. For these simulations we used currently available theoretical rmf and arf of the Wide Field Imager (WFI), including a simulated background, and assumed a modest exposure time of 50~ks.\footnote{Specifically, we used athena$\_$wfi$\_$1190$\_$onaxis$\_$w$\_$filter$\_$v20150326.rsp (rmf), athena$\_$wfi$\_$1190$\_$onaxis$\_$w$\_$filter$\_$v20150326.arf (arf) and athena$\_$wfi$\_$1469$\_$bkgd$\_$sum$\_$psf$\_$onaxis$\_$w$\_$filter$\_$20150327.pha (bg), taken from https://www.cosmos.esa.int/web/athena/resources-by-esa} The simulated spectra were rebinned into channels with a minimum of 20 counts per bin and then we performed the same analysis carried out for the \chan\ data on these synthetic \athena\ data. 

In our current work, five of the seven selected sources did not give satisfactory results mainly because of the low statistics (see Section~\ref{subsec:statistics}). Since 4U 1608--52 is a source for which different techniques can in principle be employed to lead to $M$ and $R$ constraints (see Section~\ref{subsec:4u1608}), making this a promising target, we simulated the quiescent spectrum of this particular source. Recall that the single \chan\ observation available for 4U 1608--52 (23~ks) did not provide meaningful constraints on the NS parameters (Section~\ref{subsec:statistics}), presumably due to the combined effect of a relatively high absorption column density (reducing the number of photons at low energies) and the presence of a power-law spectral component (increasing the level of degeneracy with the thermal component). 

The results of our \athena\ simulations for 4U 1608--52 are shown in Figure~\ref{fig:Athena}, where they are directly compared with the currently available \chan\ data. Leveraging the enhanced statistics of the simulated data we tried to constrain both mass and radius. Using a composite thermal and power-law model, we obtain best-fit values for the mass and radius of $M=$1.735$\pm$0.003~$\Msun$ and $R=$10.91$^{+0.14}_{-0.19}$~km. These simulations demonstrate that very accurate constraints on the mass and radius can be obtained from quiescent spectra with significantly improved statistics (at a level of less than a few percent if the distance is well known), even if the absorption column density is relatively high and if there is a power-law component present. When taking into account the full distance range estimated for \uu, an additional uncertainty of $\delta_{\mathrm{dist}} \sim 2$~km on the radius is obtained.

We also tested if the dependence on the energy range would remain an issue if better quality data is available. To this end we simulated an \athena\ spectrum for \exo\ based on the best-fit values of our fits with both mass and radius free (see Table \ref{tab:MRFinal}) in the range 0.3--10 keV, and then fitted this spectrum with the model \textsc{tbabs}$\times$\textsc{nsatmos}. We launched one fit using the energy range 0.3--10 keV and another one on the energy range 0.5--10 keV, to check if the same energy range bias observed in \chan\ (and \xmm) spectra may still be an issue dealing with \athena\ data.  The two results are perfectly consistent with the input values for the simulation and with each other:
we obtain $M=$1.641$^{+0.011}_{-0.080}~\Msun$ and $R=$10.06$^{+0.05}_{-0.06}$km when fitting over the 0.3--10 keV range and $M=$1.646$^{+0.016}_{-0.013}~\Msun$ and $R=$10.02$^{+0.11}_{-0.09}$~km when considering 0.5--10~keV instead. This suggests that the dependence on the energy range is something that may depend only on the quality of the spectral data. Due to the lack of a power-law component, the radius can be measured even more accurately than for \uu, with $\sim$1\% errors, if the distance were known accurately. Accounting for the distance uncertainty of \exo\ yields and error of $\delta_{\mathrm{dist}} \sim 2.5$~km on the NS radius.

We note that these simulations are based on current assumptions for e.g. coating efficiency, coating thickness versus annulus, roughness, etc., for the \athena/WFC design and are presented merely to illustrate the advances brought by higher quality data. The results from the simulations should of course not be taken at face value. 

\section{Summary and outlook}\label{sec:conclusions}

In this work we analyzed \chan\ spectra of seven quiescent LMXBs residing in the Galactic Plane. Our main aim was to probe to what extent such objects can be used to constrain the radius (and mass) of the NS, and in particular to test the robustness of such results against several biases this method is subject to. The motivation for this work is that field LMXBs have some advantages over globular cluster sources, mostly used for this type of analysis so far, particularly if their distances could be more accurately determined (e.g. with \gaia).

We found that for only two of the sources in our sample, \exo\ and \aql, the quality of the spectral data was sufficient to obtain some constraints on the NS radius and mass. For the other five NS LMXBs no such results could be obtained due to limited statistics ($\lesssim$ 3000 counts in a spectrum), which is in part due to factors like the interstellar absorption, and/or the presence of a power-law spectral component. 

For \exo\ and \aql, which both have multiple \chan\ quiescence observations and a relatively low $N_{\mathrm{H}}$, we found NS radii of $R < 15$ and $ R < 19$~km, respectively (90\% confidence ranges, fitting the data between 0.3--10 keV and assuming a NS mass of $1.4~\Msun$, taking into account the full available distance range). Although these results depend on the distance and the energy range for spectral fitting, we can consider these values as upper limits for the NS radius, because these are valid under the various parameter ranges that were probed. In principle for both sources the mass and radius could be constrained simultaneously from the spectral fits, albeit with large errors (see Table \ref{tab:MRFinal}).

The NS parameters can be much less well constrained for Aql X-1 than for EXO 0748--676, which is likely due to the higher statistics of the latter (because of a combination of lower absorption column density and absence of a power-law spectral component) and the greater distance uncertainty in the former. 

Using the \chan\ data of \exo, we investigated the dependence of the inferred NS parameters on the energy range over which the spectral fits are performed. For fits performed using data between 0.5--10 keV we found systematically lower radii (and masses) than for fits that extended to a lower energy of 0.3~keV. This effect was also spotted when using \xmm\ data of this source \citep[][]{Cheng2017}. This indicates that the chosen range for spectral fitting biases the obtained NS radius and mass. However, \athena\ simulations that we performed for \exo\ return no such energy dependence, indicating that this bias can possibly be lifted with increased statistics. 

Using the \chan\ data of Aql X-1, we investigated the effect of a power-law spectral component in the data. Based on the simple and often used f-test, we initially divided the 14 observations for this source as either being a pure NS atmosphere, or requiring an additional power-law emission tail. For the former sample we found a general bad agreement between data and model, witnessed by values of $\chi^2_\nu\sim$ 1.5, and the arising of on the one hand an unexpected correlation between the obtained radius and assumed mass, and on the other hand the lack of the expected correlation between the inferred radius and assumed distance. Trends in the values for $N_{\mathrm{H}}$ and the average effective temperature obtained by the fits were observed as well. All these effects disappeared when a power-law component was added to the spectra that were originally diagnosed as likely being pure NS atmospheres. This suggests that a hard emission tail is hidden in those spectra and by not modeling it, this peculiar behavior is observed. Testing the mass and distance dependence of the NS radii inferred from spectral fits may thus both serve as simple tests to see if there might be an unmodeled power-law component present in the data.

We performed \athena\ simulations for the NS LMXB \uu\ to probe the advances that may be brought by future instrumentation. For this source we could not obtain meaningful results using the existing \chan\ data, presumably due to its relatively high column density and the presence of a hard emission tail in the quiescent spectrum. We found that with highly increased statistics the presence of a power-law spectral component does not hamper obtaining accurate constraints on the NS parameters. Indeed, the mass and radius can in principle be very accurately constrained simultaneously with \athena, provided that the distance towards the source is well known. 

Despite the advances expected to be brought by future missions, an important caveat is, as discussed in Section \ref{sec:intro}, that the presence of a power-law component in the spectrum might be interpreted as the signature of residual accretion. As argued in \citet{Wijnands2015}, a power-law component that contributes $\sim$50\% to total unabsorbed 0.5--10 keV flux very likely originates from accretion, whereas a much lower power-law contribution could have a different origin. The presence of low-level accretion can hamper reliable radius measurements. For instance, low-level accretion could keep the metal abundance in the atmosphere high enough to soften the spectrum, causing an overestimate of the NS radius \citep{Rutledge2002KS}. On the other hand, if accretion is ongoing and concentrated onto a fraction of the NS, the inferred radius may be underestimating the true one. Fortunately, there are several field LMXBs (e.g. EXO 0748--676, MXB 1659--298 and KS 1731--260 studied in this work) that have spectra fitted well by NS atmosphere models without showing evidence for an additional power-law component (or any other signs of low-level accretion such as irregular variability).

Uncertainties in the source distance results in a well-known bias in the obtained radius. This problem is more severe for field LMXBs as their distances are often less well constrained than for the ones located in globular clusters. However, new and more precise distance measurements may become available in the near future from \textit{GAIA}  \citep[Global Astrometric Interferometer for Astrophysics;][]{Gaia.etal:2016a}. This ESA space observatory has been measuring, since 2013, parallax distances of millions of stars, visible in the optical band, with an unprecedented precision.\footnote{The expected precision is 1$\%$ for nearest stars to 10$\%$ for sources close to the Galactic Center, about 10 kpc from the Earth.} One in three LMXBs have a known optical counterpart \citep{Liu2007}, and more accurate distances may be obtained with \gaia\ for some field NS LMXBs. This can strongly reduce the uncertainties on their measured radii (and masses). 

The advances brought by sensitive X-ray detectors with a high collective area, such as \athena, might provide good prospects for obtaining reliable NS radius (and mass) constraints for field LMXBs using their quiescent thermal X-ray emission. Such studies are particularly promising for sources for which more accurate distances can be determined, and that have a low hydrogen column density and small or absent contribution from a hard emission tail in their quiescent spectra. Furthermore, for some of these sources other measurements may be obtained, such as independent $R$ and $M$ estimates from thermonuclear X-ray bursts, dynamical mass measurements for sources that have bright optical counterparts, or indirect constraints from the accretion flow properties (reflection features and QPOs). Some field LMXBs, e.g. Aql X-1 and \uu, meet several of these requirements.

\section*{Acknowledgements}
The authors are grateful to the anonymous referee for a thoughtful report that helped improve the clarity of this work. AM was partly supported by an "ERASMUS+ for traineeship" grant and partly by  the NewCompStar network, COST Action MP1304. AM thanks the hospitality of the Anton Pannekoek Institute for Astronomy in Amsterdam, where most of this research was carried out. ND is supported by a Vidi grant awarded by the Netherlands organization for scientific research (NWO). RW is supported by Top-1 grant from NWO. The authors acknowledge financial contribution from the agreement ASI-INAF I/037/12/0,
as well as fruitful discussion with the international team ``The disk-magnetosphere
interaction around transitional millisecond pulsars'' at the International Space
Science Institute, Bern. This work is partially supported by the HERMES Project,
financed by the Italian Space Agency (ASI) Agreement n. 2016/13 U.O



\footnotesize{
\bibliography{biblio}
}


%

\newpage

\appendix 
\section{Description of the sources}\label{appendix:src}
In this appendix we provide detail on the properties of the seven NS qLMXBs in our analyzed sample.

\subsection{KS 1731--260}\label{subsec:ks}
The first observation of this source in outburst dates back to 1989 and the detection of X-ray bursts identified it as a NS LMXB \citep{Sunyaev1989}. The source is located in the direction of the Galactic Center and distance estimates suggests that it might be close to it; \citet{Muno2000} obtained an upper limit of $D<7$~kpc based on X-ray burst analysis, while \citet{Ozel2012} placed the source at $~\approx$8 kpc by studying the distribution of stars in its direction. In the same work, the authors analyze multiple X-ray bursts to infer $R<12$~km and $M<1.8~\Msun$ as upper limits for the NS parameters.\footnote{We note that the ``touchdown'' method applied for \ks\ \citep[as well as other sources; see][for a review]{Ozel2016} has been criticized in a number of works \citep[see e.g.][for summaries of this discussion]{Heinke2013,Miller2016}. This is mainly because the X-ray bursts used in this approach occur at high accretion rates and therefore the X-ray burst emission may be contaminated by the accretion emission \citep[see][for a review]{Degenaar2018}, which can introduce biases \citep[e.g.][]{Poutanen2014,Kajava2014}.} A NS spin frequency of 524~Hz has been determined from X-ray burst oscillations \citep[1.9~ms;][]{Smith1997}, but the binary orbital period is not known. 
\ks\ has exhibited only one, $\sim$12.5-year long, historic outburst and has been in quiescence for nearly two decades now. Quiescent X-ray studies revealed a thermally-dominated spectrum and a NS surface temperature that gradually cooled over time \citep[][]{Wijnands2001_ks,Wijnands2002,Cackett2006,Cackett2010KS,Merritt2016}.

\subsection{XTE J1701--462}\label{subsec:xte}
This transient NS LMXB was discoverd when it exhibited a luminous outburst in 2006--2007 \citep{Homan2007}. Analysis of its X-ray bursts gave a measurement of the source distance of $8.8 \pm 1.3$~kpc \citep[][]{Lin2009}. The spin period of the NS or orbital period of the binary are not known, although a large disk and therefore a large orbital period are expected due to its bright outburst. After the end of its outburst, \xte\ was studied in quiescence which revealed thermal emission from the NS surface, but also a strong and variable hard emission tail, as well as occasional flares that suggest residual accretion occurring in quiescence \citep{Fridriksson2010,Fridriksson2011}. No other accretion outbursts have been detected from this source.

\subsection{MXB 1659--298}\label{subsec:mxb}
Discovered in 1976, \mxb\ was immediately identified as an LMXB harboring a NS due to the detection of X-ray bursts \citep{Lewin1976}. The source shows eclipses every $\approx$7.1 hr, representing the orbital period, which points to a high inclination and proves that it is not an ultra-compact binary \citep[][]{Cominsky1984,Cominsky1989}. The distance of the source is not firmly established: \citet{Oosterbroek2001} studied the X-ray bursts and estimated $D= 11-13$~kpc, while \citet{Galloway2008} point out two different ranges (9$\pm$2 and 12$\pm$3~kpc) depending on the H content of the material burning during the X-ray burst. Deriving an accurate distance for this source is somewhat hampered by its high inclination, which introduces systematic uncertainties \citep[][]{Galloway2008_inclination}. Rapid variability during the X-ray burst have revealed that the NS spins at 567~Hz \citep[1.67~ms;][]{Wijnands2001}. Since its discovery outburst in the 1970s, \mxb\ spent most of its time in quiescence, switching on only twice again, in 2001 and in 2015. The source has been observed multiple times in quiescence, revealing a thermally-dominated X-ray spectrum and a gradually decreasing NS temperature \citep{Wijnands2003,Wijnands2004,Cackett2006,Cackett2013}. 

\subsection{EXO 0748--676}\label{subsec:exo}
Discovered in 1985 \citep[][]{Parmar1985}, this NS LMXB remained in outburst until 2008. From analyzing its X-ray bursts, a NS spin frequency of 552 Hz \citep[]{Galloway2010} and a distance of $D=7.1 \pm 1.2$~kpc have been inferred \citep[][]{Galloway2008}. Studying the X-ray eclipses displayed by the source, a high inclination of $i\simeq 75^\circ - 83^\circ$ and an orbital period of 3.82~hr have been inferred \citep[][]{Parmar1985,Wolff2009}. 

After EXO 0748--676 returned to quiescence, it was monitored in X-rays with \chan\ and \xmm\ \citep[][]{Degenaar2009,Degenaar2011,Degenaar2014,DiazTrigo2011}, primarily to study the thermal evolution of the NS. Its mass and radius have been constrained by \citet{Cheng2017} with the quiescent method applied to \xmm\ data. Interestingly, the authors find different results whether the spectrum was fitted over a range of 0.3--10~keV or  0.5--10~keV. In particular, they find in the former case $M=2.00^{+0.07}_{-0.24}~\Msun$ and $R=11.3^{+1.3}_{-1.0}$~km, while for the second energy range they obtain $M=1.50^{+0.4}_{-1.0}~\Msun$ and $R=12.2^{+0.8}_{-3.6}$~km (at 90$\%$ confidence level).

Apart from X-ray studies, optical spectroscopic studies of the quiescent counterpart of \exo\ have put some weak constraints on the NS mass, limiting it to $1.27 \lesssim M \lesssim 2.4~\Msun$ \citep[][]{Bassa2009,MunozDarias2009}. 

\subsection{Cen X-4}\label{subsec:cenx4} 
This is the most proximate of the NSs in our sample, with a distance $D=1.2\pm 0.3$~kpc determined from the quiescent optical companion \citep{Chevalier1989}. This source was observed for the first time in 1969 during a bright ($\approx$ 20 Crab at peak), two-months lasting outburst \citep{Conner1969}. The system went in outburst a second time ten years later, in 1979 \citep[e.g.][]{Kaluzienski1980}, but it has been quiescent ever since. There is no information on the spin period of the NS, but the orbital period has been measured to be 15.1~hr \citep{Chevalier1989}, i.e. the system is not an ultra-compact binary. The main-sequence nature of the companion is confirmed by studies of the optical counterpart \citep[][]{Torres2002,Davanzo2005}. Dynamical mass measurements have been attempted but did not put meaningful constraints on the mass of the NS in Cen X-4 \citep[e.g.][]{McClintock1990,Shahbaz1993,Torres2002}.

The quiescent X-ray emission of Cen X-4 has been studied extensively with different instruments \citep[e.g.][]{Vanparadijs1987,Asai1996,Campana1997,Campana2000,Cackett2010,Cackett2013_cenx4,Bernardini2013,Chakrabarty2014,Dangelo2015}.
The X-ray spectra always clearly manifest a power-law emission component and like Aql X-1, also Cen X-4 displays clear X-ray variability in quiescence over timescales of years, days and hundreds of seconds. There is strong evidence that residual accretion occurs in quiescence \citep[][]{Cackett2010,Cackett2013_cenx4,Bernardini2013,Chakrabarty2014,Dangelo2015}. This makes it an interesting comparison with the other NSs in our sample. \citet{Cackett2010} analyzed several quiescent spectra of Cen X-4 and found that the best-fit radius for the NS is comprised in a $68\%$ confidence range of 7--12.5~km for an assumed mass of $1.5~\Msun$(using data from \emph{Chandra}, \emph{Suzaku} and \emph{XMM-Newton}).

\subsection{Aql X-1}\label{subsec:aqlx1}
This transient NS LMXB was one of the first X-ray sources to be discovered \citep[e.g.][]{Friedman1967} and has been observed in outburst numerous times since. It displays an accretion outburst about once every year \citep[e.g.][]{Kaluzienski1977,Kitamoto1993,Campana2013} and is a known X-ray burster \citep[e.g.][]{Koyama1981}. The X-ray bursts have been used to estimate a distance towards the source of 3.5--4.5~kpc \citep[depending on the composition of the ignition layer;][]{Galloway2008}, whereas the optical counterpart allows for a distance of 4--8 kpc \citep[][]{Rutledge2001,Matasanchez2017}. Coherent X-ray pulsations at a frequency of 550~Hz (1.8~ms) revealed the spin period of the NS \citep[][]{Casella2008}. These pulsations were detected only during a single $\sim$150-s long data segment (out of $\sim$1~Ms of \rxte\ data), however, and the source is not expected to be highly magnetised \citep[][]{DiSalvo2003}. The companion star has been identified as a late type star of spectral class K7 to M0, orbiting around the compact primary in $\approx$ 19 hr \citep{Callanan1999}. 

Aql X-1 has been extensively studied in quiescence using several different X-ray instruments \citep[e.g.][]{Vanparadijs1987,Verbunt1994,Rutledge2002,Campana2003,Cackett2011,Cotizelati2014,Waterhouse2016, Ootes2018}. It displays a clear thermal emission component, but its quiescent spectra often contain a power-law component too, with a contribution to the overall emission that is highly variable over relatively short timescales \citep[e.g.][]{Rutledge2002,Campana2003,Cackett2011,Cotizelati2014}. Aql X-1 is of particular interest in the context of the present work, because multiple approaches can in principle be used to measure the mass and radius of its NS: Apart from X-ray bursts and quiescent thermal emission, Aql X-1 also exhibits both mHz and kHz QPOs \citep[e.g.][]{Zhang1998,Revnivtsev2001}. Furthermore, its quiescent counterpart is in principle bright enough to obtain dynamical mass measurements \citep[][]{Matasanchez2017}.

In a recent work by \citet{Li2017}, the methods of obtaining constraints on the mass and radius from quiescent spectra and X-ray bursts (detected with \emph{RXTE}) were simultaneously applied and the results of both techniques compared (assuming different values for the distance). The analysis ellipsoidal regions in a $M-R$ diagram enclosing different ranges as the distance choice changes.
For instance, for a distance of 4 (5.5)~kpc, the obtained mass and radius regions obtained from the X-ray burst analysis roughly correspond to $M\approx$0.5--1.5~M$_\odot$ and $R\approx 9-14$~km ($\approx$ 1.5-2.5 M$_\odot$ and R$>12$ km). Using quiescent data from \emph{Chandra} and \emph{XMM-Newton}, on the other hand, the shape of resulting confidence region in $M-R$ space appears more skewed and asymmetrical, and a much less strong dependency on the distance is apparent: for 4 (5.5)~kpc, the obtained mass and radius are $M\approx$0.5--1.5~M$_\odot$ and $R\approx8-10$~km ($M\approx0.5-1.8~M$$_\odot$ and $R\approx 8-12$~km).\footnote{ \citet{Li2017} used a composite model of \textsc{nsatmos} and an additional power-law component to model the quiescent X-ray spectra.} There is only limited overlap in the confidence regions isolated by the two different methods. \citet{Li2017} propose that this could possibly due to incorrectly assuming a NS atmosphere model for the thermal emission, the lack of spin correction in the quiescent method or to an unconsidered asymmetric expansion of the photosphere during the X-ray bursts.

\subsection{\uu}\label{subsec:4u1608}
This transient NS LMXB was discovered in the 1970s \citep[][]{Tananbaum1976} and goes into outburst roughly once every $\sim$1--2 years \citep[e.g.][]{Lochner1994,Simon2004}, during which it exhibits X-ray bursts \citep[e.g.][]{Murakami1980,Gottwald1987,Fujimoto1989,Galloway2008}. The X-ray bursts have been used to estimate distance of \uu, and have been argued to be between 3.4 and 4.2 kpc based on their observed spectral evolution \citep{Ebisuzaki1987}, and between 2.9 and 4.5 kpc based on the peak fluxes of the brightest X-ray bursts \citep{Galloway2008}. With a completely different method based on the study of the interstellar extinction between the source and the observer, \citet{Guver2010} determined a wider range for the allowed distances of 3.9--7.8 kpc. A study of rapid variability detected during the X-ray bursts \citep[so-called burst oscillations; e.g.][]{Strohmayer1996} revealed that the NS spins at a high frequency ($\nu_\mathrm{spin}$) of 620 Hz \citep{Muno2001,Galloway2008}. The secondary star is likely to be an H-rich donor star, such as an F- or G-type star, as inferred from the spectrum of the optical counterpart, QX Nor \citep{Wachter2002}. Furthermore the orbital period, never measured but supposedly close to 12 hr from the period of a super-hump, renders it unlikely that the secondary star is a degenerate He-rich star, as in ultra-compact LMXBs (and therefore the NS atmosphere is likely composed of H).

What makes this NS particularly interesting is that it is possible, in theory, to apply several different techniques to estimate its radius. Firstly, from analyzing its X-ray bursts, \citet{Poutanen2014} found $R=15-21$~km for $M=1.5~\Msun$ and more generally put a lower limit on the NS radius of $R>12$~km for $M=1.0$--$2.4~\Msun$.\footnote{We note that \citet{Guver2010} obtained $R=9.3\pm0.1$~km and $M=1.74\pm0.14~\Msun$ for \uu\ by studying a different sample of bursts. However, \citet{Poutanen2014} argued that this result was likely biased by the artificial cuts in the distance and color correction factor that were made in the analysis.} Secondly, \citet{Degenaar2015} modeled the Fe-K line in the reflection spectrum and obtained an upper limit of $R<21$~km for the radius of the NS (assuming $M=1.5~\Msun$). Furthermore, \uu\ displays both kHz and mHz quasi-periodic oscillations, QPOs \citep[e.g.][]{Berger1996,Mendez1998,Jonker2000,Revnivtsev2001}; such signals have been used to place some (weak) constraints on NS radii too \citep[e.g.][]{Miller1998,Stiele2016}, although not for this particular source.

\section{Spectral fitting results}
In this appendix we provide tables with the results of various X-ray spectral fits for Aql X-1 and \exo, i.e. the two sources with the highest quality spectra and the only two for which some constraints on the NS parameters could be obtained. Since the temperature is a parameter left free to vary among the simple spectra in the samples, we report here the average value for the spectra, $\langle \log T_{\mathrm{eff}} \rangle$.

\begin{table*}
\centering
\begin{tabular}{ l l l l l l l l  l l }
\hline 
& \multicolumn{4}{c}{0.3-10 keV} & & \multicolumn{4}{c}{0.5-10 keV} \\
\cmidrule(lr){2-5} \cmidrule(lr){7-10}
{$M$} & {$R$} & {$N_{\mathrm{H}}$} & {$\langle \log T_{\mathrm{eff}} \rangle$} &{$\chi^2_\nu$ (d.o.f.)} & & {R} & {$N_{\mathrm{H}}$} & {$\langle \log T_{\mathrm{eff}} \rangle$} & {$\chi^2_\nu$ (d.o.f.)}  \\
{(M$_\odot$)} &  {(km)} & {(10$^{21}$ cm$^{-2}$)} & & & & {km} & {(10$^{21}$ cm$^{-2}$)}  \\
\hline
{1.1} & {11.3$\pm$0.6} & {0.48$\pm$0.07} & {6.24$\pm$0.01}&{1.14(654)} & &  {9.8$^{+0.6}_{-0.6}$} & {0.21$\pm$0.09} & {6.26$\pm$0.01}&{1.07(616)} \\
{1.2} &  {11.2$^{+0.6}_{-0.5}$}&{0.48$^{+0.08}_{-0.06}$} & {6.24$\pm$0.01} &{1.14(654)} & & {9.6$^{+0.7}_{-0.5}$}&{0.21$^{+0.10}_{-0.08}$} & {6.27$\pm$0.01} &{1.07(616)}  \\
{1.3} &  {11.0$^{+0.6}_{-0.5}$}&{0.49$^{+0.08}_{-0.07}$}& {6.25$\pm$0.01} & {1.14(654)} & & {9.2$^{+0.7}_{-0.5}$}&{0.20$^{+0.10}_{-0.08}$} & {6.29$^{+0.01}_{-0.02}$} &{1.07(616)}\\
{1.4} &  {10.7$^{+0.7}_{-0.5}$}&{0.50$^{+0.08}_{-0.07}$} & {6.26$\pm$0.01} & {1.14(654)} & &  {8.9$^{+0.7}_{-0.5}$}&{0.22$^{+0.09}_{-0.07}$} & {6.31$^{+0.01}_{-0.02}$} &{1.07(616)} \\
{1.5} &  {10.5$^{+0.8}_{-0.5}$}&{0.50$^{+0.07}_{-0.06}$} & {6.28$^{+0.01}_{-0.02}$} & {1.14(654)} & & {8.2$^{+0.7}_{-0.5}$}&{0.22$^{+0.09}_{-0.06}$} & {6.34$\pm$0.02} &{1.07(616)} \\
{1.6} & {10.0$^{+0.8}_{-0.5}$} & {0.51$^{+0.08}_{-0.06}$} & {6.29$^{+0.01}_{-0.02}$} & {1.14(654)}& & {8.2$^{+0.7}_{-0.5}$}&{0.29$^{+0.08}_{-0.06}$} & {6.35$^{+0.02}_{-0.03}$} &{1.07(616)}\\
{1.7} &  {9.6$^{+0.9}_{-0.8}$}&{0.53$^{+0.07}_{-0.05}$}& {6.31$^{+0.01}_{-0.02}$} & {1.14(654)}  & & {8.5$^{+0.7}_{-0.5}$}&{0.38$^{+0.07}_{-0.06}$} & {6.35$^{+0.02}_{-0.03}$} & {1.08(616)} \\
{1.8} &  {9.3$^{+0.9}_{-0.5}$}&{0.56$^{+0.07}_{-0.04}$}& {{6.34$^{+0.02}_{-0.03}$}} & {1.14(654)} & & {8.9$^{+0.8}_{-0.4}$}&{0.47$^{+0.07}_{-0.05}$} & {6.35$^{+0.02}_{-0.03}$} & {1.10(616))}\\
{1.9} & {9.5$^{+0.8}_{-0.5}$}&{0.62$^{+0.06}_{-0.04}$} & {{6.33$^{+0.02}_{-0.03}$}} &{1.15(654)} & & {9.4$^{+0.9}_{-0.5}$}&{0.56$^{+0.07}_{-0.05}$} & {6.34$^{+0.02}_{-0.03}$} &{1.12(616)}\\
{2.0}& {9.9$^{+0.9}_{-0.5}$}&{0.69$^{+0.06}_{-0.05}$}& {{6.33$^{+0.02}_{-0.03}$}} &{1.17(654)} & & {9.9$^{+0.8}_{-0.6}$}&{0.65$^{+0.07}_{-0.06}$} & {{6.33$^{+0.02}_{-0.03}$}} & {1.15(616)} \\
\hline
\end{tabular}
\caption{\textbf{EXO 0748--676}. Best-fit radii for different values of the mass, kept frozen during the fits, and best-fit parameters. The values for the distance and normalization of the \textsc{nsatmos} model were frozen to 7.1 kpc and 1, respectively.}
\label{tab:REXO}
\end{table*}

\begin{table*}
\centering
\begin{tabular}{ l l l l l l l l l l l l l}
\hline 
\multicolumn{6}{c}{Mass Variable} & & & \multicolumn{4}{c}{Distance Variable} \\
\cmidrule(lr){1-6} \cmidrule(lr){8-13}
{$M$} & {$R$} & {$N_{\mathrm{H}}$} & {$\Gamma$} & {$\langle \log T_{\mathrm{eff}} \rangle$} & {$\chi^2_\nu$ (d.o.f.)} & & {D} & {R} & {$N_{\mathrm{H}}$} & {$\Gamma$} & {$\langle \log T_{\mathrm{eff}} \rangle$} & {$\chi^2_\nu$ (d.o.f.)}  \\
{(M$_\odot$)} &  {(km)} & {(10$^{21}$ cm$^{-2}$)} & & & &  & {(kpc)} & {km} & {(10$^{21}$ cm$^{-2}$)}  \\
\hline
{1.1} &  {12.9$^{+1.9}_{-1.4}$}&{4.1$^{+0.3}_{-0.2}$}&{1.2$^{+0.5}_{-0.4}$} & {6.19$^{+0.02}_{-0.03}$} & {0.87(541)} & &  {4.0} & {9.4$^{+0.9}_{-1.4}$} & {4.2$\pm$0.3} & {1.1$\pm$0.4} & {6.23$^{+0.03}_{-0.04}$} &{0.81(342)} \\
{1.2} &  {12.5$^{+1.8}_{-1.4}$}&{4.1$^{+0.3}_{-0.2}$}& {1.1$\pm$0.4}& {6.19$^{+0.02}_{-0.03}$} & {0.87(541)} & & {4.5} & {11.5$^{+1.7}_{-1.2}$}&{4.1$^{+0.4}_{-0.3}$}& {1.1$\pm$0.4} & {6.20$^{+0.02}_{-0.03}$} & {0.87(541)} \\
{1.3} &  {12.0$^{+2.0}_{-1.4}$}&{4.1$^{+0.3}_{-0.2}$}&{1.1$^{+0.5}_{-0.3}$} & {6.20$^{+0.02}_{-0.03}$} &{0.87(541)} & & {5.0} &   {12.2$^{+1.9}_{-1.3}$}&{4.1$^{+0.3}_{-0.2}$}&{1.1$^{+0.5}_{-0.3}$} & {6.21$^{+0.02}_{-0.03}$}&{0.87(541)}\\
{1.4} &  {12.2$^{+1.9}_{-1.3}$}&{4.1$^{+0.3}_{-0.2}$}&{1.1$^{+0.5}_{-0.3}$}&  {6.21$^{+0.02}_{-0.03}$}&{0.87(541)} & &  {5.5} & {13.3$^{+1.7}_{-1.5}$}&{4.1$^{+0.4}_{-0.1}$}&{1.1$^{+0.5}_{-0.3}$}& {6.21$^{+0.01}_{-0.04}$}&{0.87(541)}\\
{1.5} & {11.8$^{+1.2}_{-0.6}$}&{4.1$^{+0.3}_{-0.2}$}&{1.1$^{+0.5}_{-0.4}$}&  {6.19$^{+0.02}_{-0.03}$} & {0.87(541)} & & {6.0} & {14.4$^{+2.6}_{-1.8}$}&{4.1$^{+0.4}_{-0.2}$}&{1.1$^{+0.5}_{-0.4}$} & {6.20$^{+0.02}_{-0.03}$}& {0.87(541)} \\
{1.6} & {12.0$^{+2.0}_{-1.4}$} &{4.2$^{+0.3}_{-0.2}$}&{1.1$^{+0.5}_{-0.4}$} & {6.21$^{+0.02}_{-0.04}$}&{0.87(541)} & & {6.5} & {15.4$^{+3.0}_{-1.9}$} &{4.1$^{+0.4}_{-0.2}$}&{1.2$^{+0.5}_{-0.4}$} & {6.20$^{+0.02}_{-0.03}$} &{0.87(541)} \\
{1.7} &  {11.0$^{+2.0}_{-1.1}$}&{4.1$^{+0.3}_{-0.2}$}&{1.1$\pm$0.4}& {6.21$^{+0.03}_{-0.04}$}&{0.87(541)} & & {7.0}&{16.4$^{+2.0}_{-1.8}$} &{4.1$^{+0.4}_{-0.2}$}&{1.1$^{+0.5}_{-0.3}$}&  {6.20$^{+0.02}_{-0.03}$} &{0.87(541)} \\
{1.8} &  {12.0$^{+2.0}_{-1.2}$}&{4.1$^{+0.3}_{-0.2}$}&{1.1$^{+0.5}_{-0.3}$}& {6.24$^{+0.03}_{-0.04}$}&{0.87(541)} & & {7.5}&{17.9$^{+2.5}_{-1.7}$} &{4.1$^{+0.4}_{-0.2}$}&{1.1$^{+0.5}_{-0.3}$} &{6.20$^{+0.02}_{-0.03}$}&{0.87(541)} \\
{1.9} &  {12.0$^{+2.0}_{-1.1}$}&{4.2$\pm$0.2}&{1.2$^{+0.4}_{-0.3}$}& {6.25$^{+0.03}_{-0.05}$}&{0.87(541)} & &{8.0}&{18.0$\pm$2.0} &{4.1$^{+0.4}_{-0.2}$}&{1.1$^{+0.5}_{-0.3}$} &{6.20$^{+0.02}_{-0.03}$}&{0.87(541)}  \\
{2.0} &  {12.0$^{+2.0}_{-1.2}$}&{4.3$\pm$0.3}&{1.2$^{+0.4}_{-0.3}$}& {6.26$^{+0.03}_{-0.06}$}&{0.87(541)}\\
\hline
\end{tabular}
\caption{\textbf{Aql X-1} full sample. Left: Best-fit radii for different values of the mass, kept frozen during the fits, and best-fit parameters. The value for the distance was frozen at 5 kpc. Right: Best-fit radii for different values of the distance, kept frozen during the fits, and best-fit parameters. The value for the mass has been kept frozen to 1.4~$M_{\odot}$. The normalization for the \textsc{nsatmos} model and the energy range for \textsc{pegpwrlw} were frozen in both series of fits to values of 1 and 0.5--10~keV, respectively. These results refer to the final, all-inclusive sample of the 14 available \chan\ observations. }
\label{RAqlX1All}
\end{table*}

\begin{table*}
\centering
\begin{tabular}{ l l l l l l l l l l l }
\hline 
\multicolumn{5}{c}{Mass Variable} & & & \multicolumn{3}{c}{Distance Variable} \\
\cmidrule(lr){1-5} \cmidrule(lr){7-11}
{$M$} & {$R$} & {$N_{\mathrm{H}}$} & {$\langle \log T_{\mathrm{eff}} \rangle$} & {$\chi^2_\nu$ (d.o.f.)} & & {D} & {R} & {$N_{\mathrm{H}}$} & {$\langle \log T_{\mathrm{eff}} \rangle$} & {$\chi^2_\nu$ (d.o.f.)}  \\
{(M$_\odot$)} &  {(km)} & {(10$^{21}$ cm$^{-2}$)} & & & & {(kpc)} & {km} & {(10$^{21}$ cm$^{-2}$)}  \\
\hline
{1.1} & {5.0$^{+1.3}_{-5.0}$} & {3.17$\pm$0.16}& {6.44$^{+0.01}_{-0.08}$} &{1.53(203)} & &  {4.0} & {7.00$^{+1.00}_{-0.60}$} & {4.20$\pm$0.20} & {6.34$^{+0.05}_{-0.06}$} & {1.68(203)} \\
{1.2} & {5.2$^{+0.7}_{-0.1}$}&{3.40$^{+0.15}_{-0.20}$} & {6.44$^{+0.01}_{-0.08}$} &{1.55(203)} & & {4.5} & {7.00$^{+0.90}_{-0.70}$}&{3.90$\pm$0.20}& {6.36$^{+0.05}_{-0.09}$} &{1.64(203)}  \\
{1.3} & {5.8$^{+1.6}_{-0.3}$}&{3.60$\pm$0.20}& {6.42$^{+0.03}_{-0.09}$} & {1.57(203)} & & {5.0} & {6.4$^{+1.0}_{-0.5}$}&{3.73$^{+0.19}_{-0.14}$}& {6.43$^{+0.05}_{-0.08}$} & {1.60(203)}\\
{1.4} &  {6.4$^{+1.0}_{-0.5}$}&{3.73$^{+0.19}_{-0.14}$}& {6.43$^{+0.05}_{-0.08}$} & {1.60(203)} & &  {5.5} & {6.20$^{+1.40}_{-0.50}$}&{3.30$\pm$0.20} & {6.45$^{+0.05}_{-0.09}$} & {1.56(203)}\\
{1.5} & {6.7$^{+1.2}_{-0.6}$}&{3.90$^{+0.19}_{-0.16}$}&{6.41$^{+0.05}_{-0.10}$}&{1.63(203)} & & {6.0} &{6.10$^{+0.67}_{-0.05}$}&{3.40$^{+0.17}_{-0.30}$} & {6.45$^{+0.03}_{-0.06}$} & {1.54(203)}\\
{1.6} & {8.1$^{+1.2}_{-1.0}$} &{4.10$^{+0.20}_{-0.19}$}& {6.36$^{+0.05}_{-0.10}$} &{1.66(203)} & & {6.5} & {6.02$^{+0.45}_{-0.01}$}&{3.20$^{+0.20}_{-0.30}$} & {6.46$^{+0.02}_{-0.07}$} & {1.53(203)} \\
{1.7} & {8.3$^{+1.4}_{-0.9}$}&{4.20$^{+0.20}_{-0.18}$}&{6.35$^{+0.05}_{-0.08}$}&{1.69(203)} & & {7.0}&{6.50$^{+0.30}_{-0.20}$} & {3.10$^{+0.20}_{-0.30}$}&{6.47$^{+0.01}_{-0.10}$}&{1.52(203} \\
{1.8} & {8.8$^{+1.3}_{-1.1}$} & {4.30$\pm$0.20}&{6.34$^{+0.05}_{-0.07}$}&{1.72(203)} & & {7.5} & {6.30$^{+0.60}_{-0.05}$} & {2.90$^{+0.20}_{-0.30}$} & {6.47$^{+0.01}_{-0.10}$} & {1.51(203)} \\
{1.9} & {9.3$^{+1.3}_{-1.1}$}&{4.40$\pm$0.20}&{6.33$^{+0.03}_{-0.05}$}&{1.75(203)} & &{8.0} & {6.50$^{+0.70}_{-0.05}$} & {2.80$^{+0.20}_{-0.30}$} & {6.47$^{+0.01}_{-0.10}$} & {1.51(203)}  \\
{2.0} &  {9.8$^{+1.6}_{-1.4}$}&{4.60$\pm$0.20}&{6.33$^{+0.04}_{-0.05}$}&{1.78(203)}\\
\hline
\end{tabular}
\caption{\textbf{Aql X-1 "No Power-Law" sample}. Left: Best-fit radii for different values of the mass, kept frozen during the fits, and best-fit parameters. The value for the distance was frozen at 5 kpc. Right: Best-fit radii for different values of the distance, kept frozen during the fits, and best-fit parameters. The value for the mass has been kept frozen to 1.4~$M_{\odot}$. The normalization for the \textsc{nsatmos} model and the energy range for \textsc{pegpwrlw} were frozen in both series of fits to values of 1 and 0.5--10~keV, respectively.}
\label{RAqlX1NPL}
\end{table*}

\begin{table*}
\centering
\begin{tabular}{ l l l l l l l l l l l l l}
\hline 
\multicolumn{6}{c}{Mass Variable} & & & \multicolumn{4}{c}{Distance Variable} \\
\cmidrule(lr){1-6} \cmidrule(lr){8-13}
{$M$} & {$R$} & {$N_{\mathrm{H}}$} & {$\Gamma$} & {$\langle \log T_{\mathrm{eff}} \rangle$} & {$\chi^2_\nu$ (d.o.f.)} & & {D} & {R} & {$N_{\mathrm{H}}$} & {$\Gamma$} & {$\langle \log T_{\mathrm{eff}} \rangle$} & {$\chi^2_\nu$ (d.o.f.)}  \\
{(M$_\odot$)} &  {(km)} & {(10$^{21}$ cm$^{-2}$)} & & & &  & {(kpc)} & {km} & {(10$^{21}$ cm$^{-2}$)}  \\
\hline
{1.1} & {12.5$^{+2.2}_{-1.3}$}&{4.1$^{+0.4}_{-0.2}$} & {1.2$\pm$0.05} & {6.19$^{+0.04}_{-0.02}$} & {0.81(342)} & &  {4.0} & {9.3$^{+1.7}_{-1.4}$}&{4.1$\pm$0.2}&{1.2$^{+0.5}_{-0.4}$}&{6.24$^{+0.04}_{-0.05}$}&{0.81(335)} \\
{1.2} & {13.2$^{+2.7}_{-1.7}$}&{4.1$^{+0.4}_{-0.3}$}&{1.3$^{+0.5}_{-0.4}$}&{6.19$^{+0.03}_{-0.04}$}&{0.81(342)} & & {4.5} & {7.0$^{+5.5}_{-0.5}$}&{4.1$^{+0.8}_{-0.2}$}&{1.2$^{-0.4}_{+0.9}$}& {6.29$\pm$0.03}  &{0.81(335)} \\
{1.3} & {12.3$^{+2.4}_{-1.2}$}&{4.1$^{+0.4}_{-0.3}$} & {1.2$\pm$0.5} & {6.21$^{+0.02}_{-0.06}$} & {0.81(342)} & & {5.0} & {12.8$^{+3.4}_{-1.1}$}&{4.2$^{+0.5}_{-0.2}$}&{1.2$^{+0.6}_{-0.3}$}&{6.21$^{+0.02}_{-0.06}$}&{0.81(335)}\\
{1.4} & {12.8$^{+3.4}_{-1.1}$}&{4.2$^{+0.5}_{-0.2}$}&{1.2$^{+0.6}_{-0.3}$}&{6.21$^{+0.02}_{-0.06}$} & {0.81(335)} & &  {5.5} & {13.2$^{+2.2}_{-1.7}$}&{4.1$\pm$0.3}&{1.2$^{+0.5}_{-0.4}$}&{6.20$^{+0.03}_{-0.04}$}&{0.81(335)}\\
{1.5} & {12.0$^{+2.6}_{-1.2}$}&{4.1$^{+0.4}_{-0.2}$}&{1.2$\pm$0.5}& {6.22$^{+0.04}_{-0.05}$} & {0.81(342)} & & {6.0} & {14.3$^{+2.4}_{-1.9}$}&{4.1$\pm$0.3}&{1.2$\pm$0.5}&{6.20$^{+0.03}_{-0.04}$}&{0.81(335)} \\
{1.6} & {11.8$^{+2.3}_{-1.6}$} &{4.1$^{+0.4}_{-0.2}$}&{1.2$^{+0.5}_{-0.4}$} & {6.23$^{+0.04}_{-0.05}$} &{0.81(342)}& & {6.5} & {15.0$^{+3.0}_{-2.0}$}&{4.1$\pm$0.3}&{1.2$^{+0.3}_{-0.4}$} & {6.20$^{+0.03}_{-0.04}$} &{0.81(335)} \\
{1.7} & {11.0$\pm$2.0} & {4.1$\pm$0.3} & {1.2$\pm$0.4} & {6.25$^{+0.05}_{-0.08}$} & {0.81(342)} &  & {7.0}&{16.0$^{+3.0}_{-2.0}$}&{4.1$^{+0.4}_{-0.3}$}&{1.2$^{+0.5}_{-0.4}$}&{6.20$^{+0.03}_{-0.04}$}&{0.81(335)} \\
{1.8} & {10.3$^{+2.2}_{-1.8}$}&{4.1$^{+0.3}_{-0.2}$}&{1.1$^{+0.5}_{-0.3}$} & {6.28$^{+0.06}_{-0.08}$}&{0.81(342)} & & {7.5}&{18.0$^{+3.0}_{-2.0}$}&{4.1$^{+0.4}_{-0.3}$}&{1.2$^{+0.6}_{-0.4}$}&{6.19$^{+0.03}_{-0.04}$}&{0.81(335)} \\
{1.9} & {9.8$^{+2.5}_{-1.9}$}&{4.1$\pm$0.2}&{1.2$^{+0.5}_{-0.4}$}&{6.30$^{+0.06}_{-0.08}$}&{0.82(342)} & &{8.0}& {18.0$^{+3.0}_{-2.0}$}&{4.0$^{+0.4}_{-0.3}$}&{1.2$^{+0.5}_{-0.4}$}&{{6.20$^{+0.03}_{-0.04}$}}&{0.81(335)}  \\
{2.0} & {10.4$^{+2.7}_{-1.8}$}&{4.2$\pm$0.2}&{1.3$\pm$0.4}&{{6.29$^{+0.06}_{-0.08}$}}&{0.81(342)}\\
\hline
\end{tabular}
\caption{\textbf{Aql X-1 "Power-Law" sample}. Left: Best-fit radii for different values of the mass, kept frozen during the fits, and best-fit parameters. The value for the distance was frozen at 5 kpc. Right: Best-fit radii for different values of the distance, kept frozen during the fits, and best-fit parameters. The value for the mass has been kept frozen to 1.4~$M_{\odot}$. The normalization for the \textsc{nsatmos} model and the energy range for \textsc{pegpwrlw} were frozen in both series of fits to values of 1 and 0.5--10~keV, respectively. }
\label{RAqlX1PL}
\end{table*}

\begin{table*}
\centering
\begin{tabular}{ l l l l l l l l l l l }
\hline 
\multicolumn{5}{c}{Mass Variable} & & & \multicolumn{3}{c}{Distance Variable} \\
\cmidrule(lr){1-5} \cmidrule(lr){7-11}
{$M$} & {$R$} & {$N_{\mathrm{H}}$} & {$\langle \log T_{\mathrm{eff}} \rangle$} & {$\chi^2_\nu$ (d.o.f.)} & & {D} & {R} & {$N_{\mathrm{H}}$} & {$\langle \log T_{\mathrm{eff}} \rangle$} & {$\chi^2_\nu$ (d.o.f.)}  \\
{(M$_\odot$)} &  {(km)} & {(10$^{21}$ cm$^{-2}$)} & & & & {(kpc)} & {km} & {(10$^{21}$ cm$^{-2}$)}  \\
\hline
{1.1} & {13.2$^{+2.2}_{-1.5}$} & {4.2$^{+0.4}_{-0.3}$} &  {6.17$^{+0.02}_{-0.03}$}&{0.98(197)} & &  {4.0} & {10.6$^{+1.9}_{-1.1}$} & {4.2$^{+0.3}_{-0.2}$}&{6.20$^{+0.03}_{-0.04}$}&{0.98(197)}  \\
{1.2} & {13.2$^{+2.2}_{-1.6}$} & {4.2$^{+0.4}_{-0.3}$} & {6.18$^{+0.02}_{-0.03}$} & {0.98(197)} & & {4.5} & {12.0$^{+2.0}_{-1.3}$} & {4.2$^{+0.4}_{-0.3}$}&{6.19$^{+0.03}_{-0.04}$}&{0.98(197)}\\
{1.3} & {13.2$^{+2.2}_{-1.5}$} & {4.2$^{+0.4}_{-0.3}$} & {6.18$^{+0.02}_{-0.03}$} & {0.98(197)} & & {5.0} & {13.6$^{+2.9}_{-1.3}$} & {4.2$^{+0.4}_{-0.3}$}& {6.18$^{+0.02}_{-0.03}$} & {0.98(197)}\\
{1.4} &  {13.6$^{+2.9}_{-1.3}$} & {4.2$^{+0.4}_{-0.3}$}& {6.18$^{+0.02}_{-0.03}$} & {0.98(197)} & &  {5.5} & {14.2$^{+2.5}_{-1.7}$} & {4.2$^{+0.4}_{-0.3}$} & 6.19$^{+0.02}_{-0.03}$ & {0.98(197)} \\
{1.5} & {13.3$^{+2.4}_{-1.4}$} &  {4.2$^{+0.4}_{-0.3}$} & {6.19$^{+0.04}_{-0.02}$} & {0.98(197)} & & {6.0} &{15.3$^{+2.5}_{-1.7}$} & {4.2$^{+0.4}_{-0.3}$} &{6.19$^{+0.03}_{-0.02}$}&{0.98(197)}\\
{1.6} &{12.7$^{+2.0}_{-1.4}$} & {4.2$^{+0.4}_{-0.3}$} & 6.20$^{+0.02}_{-0.04}$ & {0.98(197)} & & {6.5} & {16.0$^{+3.0}_{-2.0}$} & {4.2$^{+0.4}_{-0.3}$} & {6.19$^{+0.02}_{-0.03}$} &{0.98(197)} \\
{1.7} & {13.0$^{+2.0}_{-1.4}$} & {4.2$\pm$0.3}& 6.21$^{+0.03}_{-0.04}$ & {0.98(197)} & & {7.0}& {17.0$^{+3.0}_{-2.0}$} & {4.1$\pm$0.4} & {6.19$^{+0.02}_{-0.03}$} & {0.98(197)} \\
{1.8} & {12.0$^{+2.2}_{-1.4}$} &  {4.2$\pm$0.3}& 6.23$^{+0.03}_{-0.04}$ & {0.98(197)} & & {7.5} & {18.0$^{+3.0}_{-2.0}$} & {4.1$^{+0.3}_{-0.2}$} & 6.19$^{+0.02}_{-0.03}$ & {0.98(197)} \\
{1.9} & {12.7$^{+2.1}_{-1.4}$} & {4.3$\pm$0.3} & {6.22$^{+0.03}_{-0.04}$} & {0.98(197)} & & {8.0} & {19.0$^{+3.0}_{-2.0}$} & {4.1$^{+0.4}_{-0.3}$}& 6.19$^{+0.02}_{-0.03}$ &{0.98(197)} \\
{2.0} & {12.5$^{+2.2}_{-1.4}$} & {4.4$^{+0.3}_{-0.2}$} & {6.23$^{+0.03}_{-0.04}$} & {0.98(197)}\\
\hline
\end{tabular}
\caption{\textbf{Aql X-1 sample originally indicated as "No Power-Law" but now with a $\Gamma = 1.2$ power-law component added}. Left: Best-fit radii for different values of the mass, kept frozen during the fits, and best-fit parameters. The value for the distance was frozen at 5 kpc. Right: Best-fit radii for different values of the distance, kept frozen during the fits, and best-fit parameters. The value for the mass has been kept frozen to 1.4~$M_{\odot}$. The normalization for the \textsc{nsatmos} model and the energy range for \textsc{pegpwrlw} were frozen in both series of fits to values of 1 and 0.5--10~keV, respectively.}
\label{RAqlX1NPL12}
\end{table*}

\begin{table*}
\centering
\begin{tabular}{ l l l l l l }
\hline 
{$D$} & & {$R$} & {$N_{\mathrm{H}}$} & {$\langle \log T_{\mathrm{eff}} \rangle$}& {$\chi^2_\nu$ (d.o.f.)} \\
{(kpc)} & & {(km)} & {(10$^{21}$cm$^{-2}$)} \\
\hline
{6.1} & & {8.9$^{+0.7}_{-0.5}$}&{0.52$^{+0.08}_{-0.06}$} & {6.28$^{+0.01}_{-0.02}$} &{1.14(654)} \\
{6.6} & & {9.6$^{+0.8}_{-0.5}$}&{0.49$^{+0.09}_{-0.06}$} & {6.27$^{+0.01}_{-0.02}$} &{1.14(654)} \\
{7.1} & & {10.7$^{+0.7}_{-0.5}$}&{0.50$^{+0.08}_{-0.07}$} & {6.26$\pm$0.01}  & {1.14(654)} \\
{7.6} & &{11.5$^{+0.8}_{-0.6}$}&{0.48$^{+0.08}_{-0.06}$} & {6.25$\pm$0.01} & {1.14(654)} \\
{8.1} &  & {12.4$^{+0.7}_{-0.6}$}&{0.48$^{+0.08}_{-0.07}$} & {6.25$\pm$0.01} & {1.14(654)} \\
\hline
\end{tabular}
\caption{\textbf{EXO 0748--676}. Best-fit radii for different values of the distance, kept frozen during the fits, and best-fit parameters. The values for the mass and normalization of the \textsc{nsatmos} model were frozen to 1.4 $M_\odot$ and 1, respectively.}
\label{tab:REXODist}
\end{table*}

\end{document}